\documentclass{mn2e}
\usepackage{epsfig}
\usepackage{aasj}

\renewcommand{\vec}[1]{\mbox{\boldmath $\displaystyle #1$}}
\newcommand{\grad}{\vec{\nabla}}

\newcommand{\vdot}{\vec{\cdot}}
\newcommand{\vcross}{\vec{\times}}

\newcommand{\curl}{\grad\vcross\,}
\newcommand{\avZ}{\left<Z\right>}

\newcommand{\gtrsim}{\ga}
\newcommand{\lesssim}{\la}
\newcommand{\bi}{\bibitem[]{}}

\title[Magnetic Field Evolution in Accreting White Dwarfs]{Magnetic
Field Evolution in Accreting White Dwarfs}
\author[A. Cumming]{Andrew Cumming\thanks{Hubble Fellow}\\Department of Astronomy
and Astrophysics, University of California, Santa Cruz, CA 95064, USA\\email: cumming@ucolick.org}

\begin{document}

\maketitle

\begin{abstract}
We discuss the evolution of the magnetic field of an accreting white
dwarf. We calculate the ohmic decay modes for accreting white dwarfs,
whose interiors are maintained in a liquid state by compressional
heating. We show that the lowest order ohmic decay time is $(8$--$12)$
billion years for a dipole field, and $(4$--$6)$ billion years for a
quadrupole field. We then compare the timescales for ohmic diffusion
and accretion at different depths in the star, and for a simplified
field structure and assuming spherical accretion, study the
time-dependent evolution of the global magnetic field at different
accretion rates. We neglect mass loss by classical nova explosions and
assume the white dwarf mass increases with time. In this case, the
field structure in the outer layers of the white dwarf is
significantly modified for accretion rates above the critical rate
$\dot M_c\approx (1$--$5)\times 10^{-10}\ M_\odot\ {\rm yr^{-1}}$. We
consider the implications of our results for observed systems. We
propose that accretion-induced magnetic field changes are the missing
evolutionary link between AM Her systems and intermediate polars. The
shorter ohmic decay time for accreting white dwarfs provides a partial
explanation of the lack of accreting systems with $\approx 10^9\ {\rm
G}$ fields. In rapidly accreting systems such as supersoft X-ray
sources, amplification of internal fields by compression may be
important for Type Ia supernova ignition and explosion. Finally,
spreading matter in the polar cap may induce complexity in the surface
magnetic field, and explain why the more strongly accreting pole in AM
Her systems has a weaker field. We conclude with speculations about
the field evolution when classical nova explosions cause the white
dwarf mass to decrease with time.
\end{abstract}


\section{Introduction}\label{sec:Intro}

There are now about 65 isolated magnetic white dwarfs known, and
a further 90 accreting magnetic white dwarfs (see Wickramasinghe \&
Ferrario 2000, hereafter WF, for a recent review). Interesting
differences are emerging between the isolated and accreting
populations. Measured field strengths of the isolated WDs range from
$3\times 10^4\ {\rm G}$ to $10^9\ {\rm G}$, whereas the range of field
strengths of accreting WDs may be smaller. The magnetic field
strengths in the AM Her binaries are directly measured to be in the
range $10^7$--$2\times 10^8\ {\rm G}$, and the magnetic fields of the
intermediate polars (IPs) are inferred to range from $10^7\ {\rm G}$
down to $\approx 10^5\ {\rm G}$. The observed fraction of magnetic
systems is 5\% for isolated WDs, but 25\% for accreting systems. There
is evidence that isolated magnetic white dwarfs are more massive than
the typical $0.6\ M_\odot$ non-magnetic WD, with a mean mass $\gtrsim
0.95\ M_\odot$. Measurements of accreting magnetic white dwarfs,
however, give masses $\approx 0.7\ M_\odot$, consistent with measured
masses of non-magnetic ($B\lesssim 10^5\ {\rm G}$) accretors. One
similarity between isolated and accreting WDs is that the magnetic
field is often seen to be complex, with a significant quadrupole (or
offset dipole) component.

The origin and evolution of white dwarf magnetism is an important, but
as yet unsolved, part of our understanding of stellar
evolution. Several papers have discussed the evolution of magnetic
fields in isolated, cooling white dwarfs. Ohmic decay of magnetic
fields was first addressed by Chanmugam \& Gabriel (1972), and
Fontaine, Thomas, \& van Horn (1973), who showed that the decay time
of the lowest order decay eigenmodes was longer than the evolutionary
time of the white dwarf. Wendell, van Horn, \& Sargent (1987;
hereafter WVS) followed the evolution of different decay modes along a
cooling sequence. They found that the fundamental dipole mode decayed
by a factor of two in $10^{10}$ years. These studies suggested that WD
magnetic fields are fossil fields left over from previous
evolution. For example, one proposal consistent with the high masses
($\gtrsim 0.95\ M_\odot$) of magnetic white dwarfs is that they result
from evolution of the strongly magnetic ($10^2$--$10^4\ {\rm G}$) Ap
and Bp main sequence stars. The higher order field components decay
more rapidly, however, leading Muslimov, van Horn, \& Wood (1995,
hereafter MVW) to study the Hall effect as a way to generate field
complexity during the lifetime of the white dwarf.

The result that white dwarf magnetic fields evolve only very slightly
over their lifetime has been applied in almost all studies of
accreting white dwarfs. For example, in evolutionary studies of
magnetic cataclysmic variables, it is presumed that the magnetic field
of the white dwarf does not change with time. One exception is King
(1985), who applied the results of Moss (1979) to accreting systems,
and suggested that meridional currents could submerge magnetic flux in
the outer layers of rotating magnetic white dwarfs. It is the purpose
of this paper to show that the process of accretion itself may
significantly change the surface magnetic field of an accreting white
dwarf.
 
Many authors have studied the effects of accretion on the magnetic
field of accreting neutron stars. This is motivated by the idea that
the rapidly rotating millisecond radio pulsars are produced by
accretion, which spins up the neutron star to short periods, and
perhaps causes a reduction in magnetic field strength from the
$10^{12}\ {\rm G}$ fields seen in most radio pulsars to the
$10^8$--$10^9\ {\rm G}$ fields of the millisecond pulsars (see
Bhattacharya 1995 for a review).  In a recent paper, Cumming, Zweibel,
\& Bildsten (2001; hereafter CZB) returned to the suggestion that the
accretion flow directly screens the internal magnetic field (Romani
1990, 1995).  They compared the timescales for ohmic diffusion and
accretion in the thin outer layers of the neutron star, and computed
steady-state magnetic profiles, taking the field to be horizontal and
the accreted matter to be unmagnetized. For this simplified geometry,
the field was found to be strongly screened for accretion rates
greater than the critical rate $\dot M\sim 10^{-10}\ M_\odot\ {\rm
yr^{-1}}$, whereas at lower accretion rates, the underlying magnetic
field was able to penetrate the freshly accreted material. This result
fits nicely with the observation that the only weakly-magnetic
accreting neutron star to show its magnetic field directly (the
accreting millisecond X-ray pulsar, SAX~J$1808.4$-$3658$; Wijnands \&
van der Klis 1998; Chakrabarty \& Morgan 1998) has a lower accretion
rate than other systems (time-averaged $\dot M\approx 10^{-11}\
M_\odot\ {\rm yr^{-1}}$), low enough that screening would not be
effective in hiding its field.

A crucial question for the evolution of accreting white dwarfs is
whether the white dwarf mass increases or decreases with time. For
accretion rates $\lesssim 10^{-7}\ M_\odot\ {\rm yr^{-1}}$, the
accreted hydrogen and helium burns unstably, giving rise to classical
nova explosions (e.g., see Fujimoto 1982; MacDonald 1983). The
observed heavy element enrichment of nova ejecta has been used to
argue that the nova explosion excavates material from the white dwarf,
so that its mass is decreasing with time (e.g., Livio \& Truran
1992). However, this is a somewhat open question, both because of
uncertainty in measurements of ejecta masses, and in theoretical
modelling. Theoretical calculations find that mixing of extra CNO
nuclei into the burning shell is needed to achieve rapid enough energy
release to drive novae with rapid rise times. However, the mixing
mechanism is presently unknown. Prialnik and Kovetz (1995) addressed
this question with hydrodynamic simulations including diffusion, and
found that the white dwarf mass decreased for $\dot M<10^{-9}\
M_\odot\ {\rm yr^{-1}}$, and increased for $\dot M>10^{-7}\ M_\odot\
{\rm yr^{-1}}$. At intermediate accretion rates, the mass loss
depended on the white dwarf temperature. The accretion rates of IPs
are estimated to be $\sim 10^{-9}\ M_\odot\ {\rm yr^{-1}}$ (see
discussion in \S 5.1), and their core temperatures in the middle of
the range considered by Prialnik and Kovetz (1995) (see \S 3.2). On
the other hand, AM Her systems accrete at typical rates $\approx
5\times 10^{-11}\ M_\odot\ {\rm yr^{-1}}$ (\S 5.1). Thus a range of
behaviour might be expected within the population of accreting
magnetic white dwarfs. The magnetic field of the white dwarf itself
may affect the nova outburst, for example by enhancing mass loss in
the equatorial plane (Livio, Shankar, \& Truran 1988; Livio 1995).

In this paper, we assume that the white dwarf mass {\it increases}
with time during accretion, and study the consequences for the
evolution of the white dwarf magnetic field. We leave the case of
decreasing white dwarf mass for a future study. We first show that
comparing accretion and ohmic diffusion times in white dwarfs gives
the same critical accretion rate $\dot M_c\approx (1$--$5)\times
10^{-10}\ M_\odot\ {\rm yr^{-1}}$ as found for neutron stars by CZB
(for reasons we describe in \S 3), and then go on to calculate the
evolution of the magnetic field. We show that the resulting field
evolution may explain several properties of observed systems.

A complication in calculating the evolution of the magnetic field in
both accreting neutron stars and white dwarfs is that the accretion
flow is channelled onto the magnetic polar cap (for white dwarfs, this
occurs for $B\gtrsim 10^5\ {\rm G}$). Rather than tackle the complex
problem of the subsequent spreading of matter and evolution of the
magnetic field, CZB presumed that the neutron star magnetic field was
buried by the accretion flow into a flattened configuration, and then
used a plane-parallel model to ask what accretion rate was needed to
keep the field from reemerging by ohmic diffusion. In this paper, the
simplification we make is to study the evolution of the global
magnetic field {\it under spherical accretion}. The accreted matter
is expected to spread away from the polar cap when the magnetic
tension force no longer supports the hydrostatic pressure (e.g.,
Hameury et al.~1983), so that spherical accretion is a good
approximation except in a thin layer of mass $\Delta M\lesssim
10^{-10}\ M_\odot\ B_7^2$ for a $0.6 M_\odot$ white dwarf, where
$B_7=B/10^7\ {\rm G}$ is the surface field strength. We include the
uncertainty of what happens in this thin spreading layer as an
uncertainty in the surface boundary condition for the spherically
accreting models (see discussion in \S 4.1).

We start in \S 2 by considering ohmic decay in accreting white dwarfs,
which have liquid interiors because of compressional heating by
accretion. In \S 3, we compare the timescales for ohmic decay and
accretion, and relate our results to those of CZB. In \S 4, we
calculate the global evolution of the field, under simplifying
assumptions about its geometry, for different accretion rates. We
relate our models to the recent calculations of Choudhuri \& Konar
(2002) for accreting neutron stars, which adopt a similar approach. We
find that for $\dot M>\dot M_c$, the surface field strength is reduced
by accretion because ohmic diffusion is not rapid enough to allow
penetration of field into the newly accreted layers. In \S 5 we
discuss the implications of our results for observed systems. In \S 6,
we conclude with some speculations about magnetic field evolution in
the alternative scenario of decreasing white dwarf mass with time. In
the Appendix, we describe our calculations of electrical conductivity
for arbitrary degeneracy.

\section{Ohmic Decay in Liquid White Dwarfs}

In this section, we discuss ohmic decay of the magnetic field,
neglecting the effects of accretion for now. In an accreting white
dwarf, compressional heating of the core results in a central
temperature $\gtrsim 10^7\ {\rm K}$ (Nomoto 1982), so that the
interior is kept in a liquid state. In that case, the electrical
conductivity is independent of temperature, allowing us to use simple
zero-temperature white dwarf models. We calculate the ohmic decay
modes following the original work of Chanmugam \& Gabriel (1972),
Fontaine et al.~(1973), and WVS. We show that the lowest order decay
mode has a decay time $\approx (8-12)\times 10^{9}\ {\rm yrs}$,
depending on white dwarf mass.

\subsection{Electrical conductivity and decay time}

Ohmic decay is described by the induction equation
\begin{equation}\label{eq:decay}
{\partial\vec{B}\over\partial t}=-\curl\left(\eta\curl\vec{B}\right),
\end{equation}
where $\eta=c^2/4\pi\sigma$ is the magnetic diffusivity, and $\sigma$
the electrical conductivity. The associated characteristic decay
timescale is $\approx 4\pi\sigma L^2/c^2$, where $L$ the
lengthscale over which $\vec{B}$ changes.

Compressional heating maintains the interior of most accreting white
dwarfs in a liquid state. The melting temperature $T_{\rm melt}$ is
determined by the condition $\Gamma\approx 173$ (e.g., Farouki \&
Hamaguchi 1993), where $\Gamma=(Ze)^2/k_BTa$ measures the ratio of
Coulomb to thermal energy. For simplicity, we write $\Gamma$ for a
single species of ion with charge $Z$, mass $A$ and interion spacing
$a$, giving
\begin{equation}
T_{\rm melt}\approx 3\times 10^6\ {\rm K}\ \rho_6^{1/3}\left({Z\over
7}\right)^{5/3}\left({2Z\over A}\right).
\end{equation}
Compressional heating gives core temperatures $\gtrsim 10^7\ {\rm
K}>T_{\rm melt}$ (Nomoto 1982), thus we expect most accreting white
dwarfs to have liquid interiors. It is possible that some massive
white dwarfs with central densities $\gtrsim 10^8\ {\rm g\ cm^{-3}}$
accreting at low enough rates have solid cores, but we do not consider
these systems here.

The electrical conductivity in the liquid interior is set by
collisions between the degenerate electrons and the non-degenerate
ions. The conductivity is
\begin{equation}\label{eq:sig}
\sigma={8.5\times 10^{21}\ {\rm s^{-1}}\over \Lambda_{ei}\avZ} {x^3\over 1+x^2},
\end{equation}
(e.g.~Yakovlev \& Urpin 1980; Itoh et al.~1983; Schatz et al.~1999),
where $x=p_F/m_ec$ measures the Fermi momentum $p_F$ of the electrons,
$\Lambda_{ei}\approx 1$ is the Coulomb logarithm, and $\avZ$ is a
measure of the mean nuclear charge, given by $\avZ=\mu_e\sum
X_iZ_i^2/A_i$, with $Z_i$ the nuclear charge, $A_i$ the mass, and
$X_i$ the mass fraction of species $i$, and $\mu_e$ the mean molecular
weight per electron.  In terms of $x$, the density is $\rho=1.9\times
10^6\ {\rm g\ cm^{-3}}\ x^3\ (\mu_e/2)$.

The fact that conductivity is independent of temperature allows us to
adopt simple zero-temperature white dwarf models. In the outer
envelope, where the electrons are non-degenerate, the conductivity is
temperature-dependent. In the Appendix, we describe a method for
calculating the electrical conductivity for arbitrary
degeneracy. However, the results of this section are insensitive to
the conductivity in the outermost layers. We adopt pressure as the
independent variable, and integrate the equations of hydrostatic
balance and mass conservation outwards from the centre of the white
dwarf for an initial choice of central density. We stop the
integration once the pressure falls below some fixed fraction of the
central pressure, at which point the radius and mass are
well-determined. We assume the composition is an equal mixture by mass
of carbon and oxygen, giving $\mu_e=2$ and $\avZ=7$.

\begin{figure}
\begin{center}
\epsfig{file=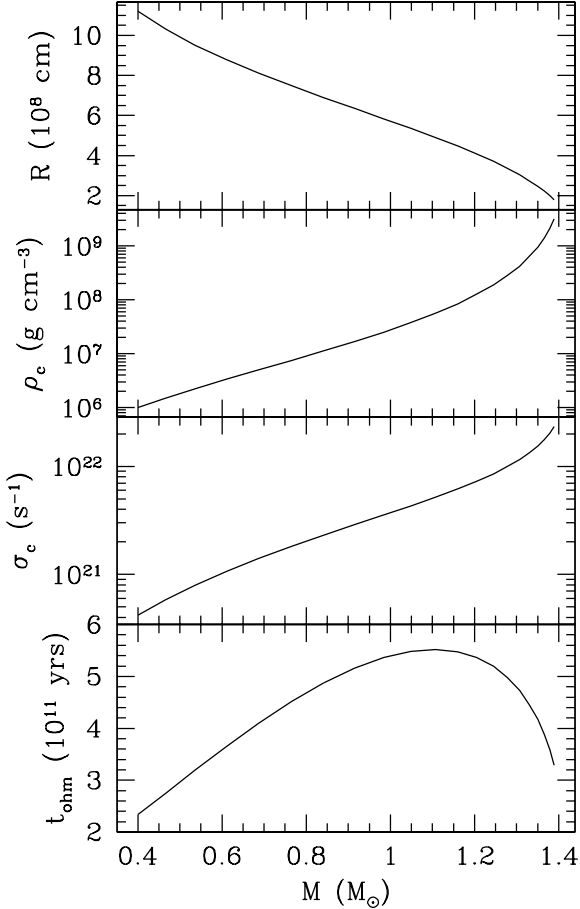,width=8.5 cm}
\end{center}
\caption{ Properties of the zero-temperature white dwarf models. We
show the radius, central density, conductivity, and ohmic time $t_{\rm
ohm}=4\pi R^2\sigma_c/c^2$ as a function of mass. The conductivity is
calculated assuming a liquid interior.
\label{fig:models}}
\end{figure}

Some properties of these zero-temperature white dwarfs are shown in
Figure \ref{fig:models}. We show the radius, central density, central
conductivity $\sigma_c$, and ohmic time at the centre, $t_{\rm
ohm}=4\pi R^2\sigma_c/c^2$, as a function of white dwarf mass.

Despite the large variations in $R$ and $\sigma_c$, the ohmic time at
the centre does not depend very strongly on white dwarf mass, ranging
from $(2$--$6)\times 10^{11}\ {\rm yrs}$. To see why, we note that at
the centre of the white dwarf $x\gtrsim 1$, so that $\sigma\propto
x\propto \rho^{1/3}$. However, if the mass-radius relation is
$R\propto M^{-1/3}$, then $\rho\propto M^2$, giving $\sigma\propto
M^{2/3}$. The ohmic time $t_{\rm ohm}\propto R^2\sigma$ is thus
approximately constant with mass. As the white dwarf mass increases,
the increase in conductivity is offset by the decreasing radius. Most
of the variation in $t_{\rm ohm}$ seen in Figure \ref{fig:models}
comes from the Coulomb logarithm (see Appendix for our calculation of
$\Lambda_{ei}$).

The difference between ohmic decay times in isolated white dwarfs with
liquid and solid cores has been pointed out previously, particularly
by MVW. For a solid, the conductivity is set by collisions of
electrons with phonons, which become more common with increasing
temperature (for example, see Yakovlev \& Urpin 1980; Baiko \&
Yakovlev 1995). An example is Figure 2 of WVS, in which both the
conductivity and decay timescale are roughly constant for the first
$\approx 3\times 10^8$ years of cooling history when the core is
liquid, but then increase with time as the core becomes solid and
cools.

\begin{figure}
\begin{center}
\epsfig{file=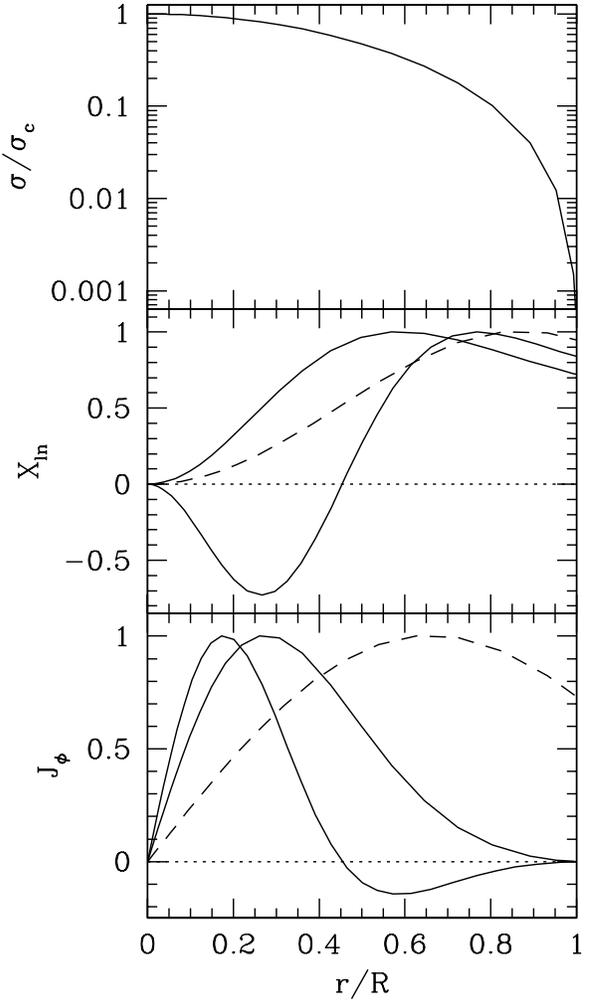, width=8.5 cm}
\end{center}
\caption{Ohmic decay modes for a $0.6 M_\odot$ white dwarf. (i) In
the upper panel, we show the conductivity as a function of radius. The
central value of conductivity in this model is $\approx 10^{21}\ {\rm
s^{-1}}$, dropping to close to the Spitzer value $\approx 5\times
10^{17}\ {\rm s^{-1}}$ (eq.~[\ref{eq:spitzer}]) at the outer
boundary. (ii) In the middle panel we show the $n=1$ and $2$ dipole
($l=1$) decay modes. We normalize each eigenfunction so that its
maximum value is unity. The $n$th mode has $(n-1)$ nodes. The dashed
curve shows the $n=1$ mode for constant conductivity with
radius. (iii) The lower panel shows the distribution of current
density for the eigenmodes shown in the middle panel.\label{fig:modes}}
\end{figure}

\subsection{Ohmic decay modes}

The ohmic time at the centre $t_{\rm ohm}$ is an overestimate of the
time for magnetic field decay, since both the conductivity and radial
lengthscale are largest at the centre. To do better, we now find the
ohmic decay eigenmodes for liquid white dwarfs, following the original
work of Chanmugam \& Gabriel (1972), Fontaine et al.~(1973), and WVS.

We assume an axisymmetric poloidal magnetic field,
$\vec{B}(r,\theta,t)=B_r(r,\theta,t)\hat{e}_r+B_\theta(r,\theta,t)\hat{e}_\theta$
which we write in terms of a vector potential $\vec{B}=\curl\vec{A}$
where $\vec{A}=A_\phi(r,\theta, t)\hat{e}_\phi$ (see Mestel 1999 for a
useful discussion). Using the identity
$\curl(\hat{e}_\phi/r\sin\theta)=0$, we find
\begin{equation}
\vec{B}={\hat{e}_\phi\over
r\sin\theta}\vcross\grad\left(rA_\phi\sin\theta\right),
\end{equation}
so that the quantity $rA_\phi\sin\theta$ labels the magnetic field
lines.

Rewriting equation (\ref{eq:decay}), we find the time evolution of
$\vec{A}$ is given by
\begin{equation}\label{eq:dAdt}
{\partial \vec{A}\over\partial t}=-\eta\curl\curl\vec{A}.
\end{equation}
We now seperate $A_\phi$ into radial and angular pieces,
\begin{equation}\label{eq:expand}
A_\phi(r,\theta,t)=\sum_l {R_l(r,t)\over r} P^1_l(\cos\theta),
\end{equation}
where $P^1_l$ is the associated Legendre function of order $1$. The
magnetic field components are
\begin{equation}\label{eq:Br}
B_r=\sum_l {l(l+1)\over r^2}R_l(r, t)P_l(\cos\theta)
\end{equation}
and
\begin{equation}\label{eq:Bt}
B_\theta=-\sum_l {1\over r}{\partial R_l(r, t)\over\partial
r}P^1_l(\cos\theta),
\end{equation}
where $P_l$ is the Legendre polynomial. We use spherically symmetric
white dwarf models\footnote{Except in the very outermost layers, it is
always a good approximation that the hydrostatic pressure $P$ is much
larger than the magnetic pressure $B^2/8\pi$.} in which the
diffusivity depends on radius $r$ only, and not on latitude
$\theta$. In this case, substituting equation (\ref{eq:expand}) into
equation (\ref{eq:dAdt}) gives
\begin{equation}\label{eq:11}
{\partial R_l\over\partial t}=\eta\left[{\partial^2R_l\over\partial
r^2}-{l(l+1)R_l\over r^2}\right],
\end{equation}
a diffusion equation for the radial function $R_l$.

Neglecting mass and density changes due to accretion, the conductivity
of a liquid white dwarf is independent of time, and depends only on
position, $\eta(r)$. In this case, we follow WVS, and look for a
solution which decays exponentially with decay time $\tau_{ln}$,
\begin{equation}
R_l(r,t)=\sum_n C_{ln} X_{ln}(r) \exp\left(-t/\tau_{ln}\right),
\end{equation}
where $C_{ln}$ is a constant giving the contribution of mode $n$, and
the differential equation
\begin{equation}\label{eq:mode}
{d^2X_{ln}\over dr^2}-{l(l+1)X_{ln}\over r^2}+{X_{ln}\over
\eta(r)\tau_{ln}}=0
\end{equation}
determines the eigenfunction $X_{ln}(r)$.

We adopt the following boundary conditions (see discussion in \S 5.6
of Mestel 1999). Inspection of equation (\ref{eq:mode}) shows that a
finite solution at the centre of the star must have $X_{ln}\propto
r^{l+1}$, giving
\begin{equation}\label{eq:bc1}
{dX_{ln}\over dr}={(l+1)X_{ln}\over r}\hspace{1cm}r\rightarrow 0.
\end{equation}
Outside the star, where the current $\vec{J}\propto\curl\vec{B}=0$, we
choose the solution which remains finite for large $r$, $R_{ln}\propto
r^{-l}$. Thus, we take
\begin{equation}\label{eq:bc2}
{dX_{ln}\over dr}=-{l X_{ln}\over r}\hspace{1cm}r\rightarrow R
\end{equation}
at the surface.

\begin{table}
\caption{Ohmic decay times ($\tau_{ln}$ in units of $10^9$
years)\label{tab:times}}
\begin{tabular}{lllllll}
& \multicolumn{3}{c}{$M=0.6\ M_\odot$} & 
\multicolumn{3}{c}{$M=1.0\ M_\odot$}\\
& \multicolumn{3}{c}{($t_{\rm ohm}=3.4\times 10^{11}$ yrs)}
& \multicolumn{3}{c}{($t_{\rm ohm}=5.4\times 10^{11}$ yrs)}\\
& $n=1$ & $n=2$ & $n=3$ & $n=1$ & $n=2$ & $n=3$\\
$l=1$& 7.6 & 2.5 & 1.2 & 12 & 4.1 & 2.0\\
$l=2$& 3.5 & 1.6 & 0.87 & 5.7 & 2.5 & 1.4\\
$l=3$& 2.0 & 1.1 & 0.66 & 3.3 & 1.7 & 1.1\\
\end{tabular}
\end{table}

Solution of equation (\ref{eq:mode}) requires the conductivity as a
function of radius. This is shown in the top panel of Figure
\ref{fig:modes} for a $0.6 M_\odot$ white dwarf, normalized to the
central value, $\sigma_c=9.6\times 10^{20}\ {\rm s^{-1}}$. The
resulting $n=1$ and $n=2$ dipole ($l=1$) decay modes $X_{ln}$ are
shown as solid lines in the middle panel of Figure
\ref{fig:modes}. The lower panel shows the current density
$J=(c/4\pi)\curl\vec{B}$. Using equations (5), (6), and (10), this may
be written in terms of $X_{ln}$ as $\vec{J}=J_\phi\ \hat{e}_\phi$,
where
\begin{equation}
J_\phi={\sigma\over c\tau_{ln}}\left({X_{ln}\over
r}\right).
\end{equation}
Note that the $n$th mode has $n-1$ nodes, the current reversing
direction at a node.

Table \ref{tab:times} gives the decay time for modes with $l=1,2$ and
$3$, and $n=1,2,$ and $3$ for $0.6\ M_\odot$ and $1.0\ M_\odot$ white
dwarfs. For both white dwarf masses, we find that the lowest order
decay mode ($l=1$ and $n=1$) has a decay time $\tau_{11}\approx t_{\rm
ohm}/40$, roughly a factor of four smaller than the analytic result
for $\eta$ independent of radius (in which case $\tau_{11}=t_{\rm
ohm}/\pi^2$, see WVS eq.~[10]).

The decay times in Table \ref{tab:times} agree well with the
calculations of WVS for liquid white dwarfs (compare their Figure
2). However, we do not agree with the later calculations of MVW, who
report decay times an order of magnitude smaller, $(3-10)\times 10^8\
{\rm yrs}$, for liquid white dwarfs (see their \S 3.3), despite using
the same conductive opacities for the liquid interior as WVS (see
their \S 2.3). The origin of this difference is not clear. The central
conductivity quoted by MVW for a $0.6\ M_\odot$ white dwarf ($\approx
2\times 10^{21}\ {\rm s^{-1}}$) agrees to a factor of two with the
value in both our models and those of WVS. In addition, changing the
conductivity profile seems unlikely to give an order of magnitude
change in the decay time. The difference may be due to an incorrect
normalization used by MVW (H.~van Horn, private communication).

To test the influence of the conductivity profile, we compared our
eigenfunctions with those of WVS and MVW (Figure 3, left panel, of
WVS; Figure 1, top panel of MVW). The agreement is good for times
$\gtrsim 10^9\ {\rm yrs}$, but differs for earlier times, for which
the WVS and MVW eigenfunctions peak at larger radius ($r/R\approx 0.8$
rather than $r/R\approx 0.6$). We have been unable to identify the
reason for this difference. The peak at larger radius likely reflects
a shallower conductivity profile. For example, in the extreme case of
constant conductivity with radius, the currents are less
centrally-concentrated than in realistic models. The reason for this
is that the currents flow in the regions of high conductivity; this
region is more extended in the constant conductivity model. The dashed
lines in the middle and lower panels of Figure \ref{fig:modes} show
the $n=1$, $l=1$ eigenmode for constant conductivity (in this case
$X_{11}\propto rj_1(r)$, where $j_1$ is the spherical Bessel function
of degree 1, see WVS). As noted above, the lowest order decay time for
constant conductivity is a factor $\approx 4$ longer than the
realistic white dwarf models.

In summary, we find that the ohmic decay time lies in the range
$\approx (8$--$12)$ billion years for a dipole field, and $\approx
(4$--$6)$ billion years for a quadrupole field, depending only
slightly on white dwarf mass.

\section{Accretion vs. Diffusion Times in the Envelope}

The long timescale for ohmic decay raises the question of the effect
of the accretion flow on the magnetic field structure. The decay
timescale of $\approx 10^{10}\ {\rm yrs}$ is the time to accrete the
whole star at $10^{-10}\ M_\odot\ {\rm yr^{-1}}$. For more rapid
accretion than this, one might expect the magnetic field structure to
be significantly altered as the magnetic field is advected by the
accretion flow. 

In fact, the evolution of the magnetic field depends on the local
accretion and diffusion timescales at different depths in the white
dwarf envelope. In this section, we compute these timescales, before
moving on to simple models of the global field structure in \S 4.

\subsection{Analytic estimates}\label{sec:analytic}

We consider the outer layers of the star, where the gravity $g=GM/R^2$
is constant. We define the column depth $y=-\int \rho dz=\Delta M/4\pi
R^2$, where $\Delta M$ is the mass above column depth $y$. Hydrostatic
balance $dP/dz=-\rho g$ becomes $dP/dy=g$, giving $P=gy$. The time to
accrete the matter above a given column depth is $t_{\rm accr}=y/\dot
m$, where $\dot m=\dot M/4\pi R^2$ is the local accretion rate per
unit area. We compare this with the ohmic diffusion time across a
scale height, $t_{\rm diff}=4\pi\sigma H^2/c^2$.

When the electrons are degenerate, the conductivity is given by
equation (\ref{eq:sig}). In the outer layers, the electrons are
non-relativistic, so that $P=6.8\times 10^{20}\ {\rm erg\ cm^{-3}}\
\rho_5^{5/3}(2/\mu_e)^{5/3}$. The pressure scale height is $H=P/\rho
g=6.8\times 10^7\ {\rm cm}\ (\rho_5^{2/3}/g_8)(2/\mu_e)^{5/3}$, giving
\begin{equation}
t_{\rm diff}=8.9\times 10^8\ {\rm yrs}\ {\rho_5^{7/3}\over
\Lambda_{ei}\avZ g_8^2}\left({2\over\mu_e}\right)^{13/3}.
\end{equation}
To evaluate the accretion timescale, we write the local accretion rate
$\dot m=\dot m_{-3}10^{-3}\ {\rm g\ cm^{-2}\ s^{-1}}$ in terms of the
global rate $\dot M=\dot M_{-10}10^{-10}\ M_\odot\ {\rm yr^{-1}}$, as
$\dot m=\dot M/4\pi R^2$, giving
\begin{equation}\label{eq:2}
\dot m_{-3}=0.51\ {\dot M_{-10}\over R_9^2}.
\end{equation}
If accretion is not occuring over the whole surface of the white
dwarf, for example magnetically-controlled accretion onto the polar
caps, the relevant quantity is the local accretion rate; we work in
terms of the global rate only for convenience. We find
\begin{equation}
t_{\rm accr}=4.3\times 10^{8}\ {\rm yrs}\ {\rho_5^{5/3}R_9^2\over
g_8\dot M_{-10}}\left({2\over\mu_e}\right)^{5/3}.
\end{equation}
The ratio of $t_{\rm diff}$ to $t_{\rm accr}$ depends on the quantity
$gR^2$, which we rewrite in terms of the mass of the star,
$g_8R_9^2=1.3(M/M_\odot)$. This gives
\begin{equation}\label{eq:rat1}
{t_{\rm diff}\over t_{\rm accr}}=1.6\ {\rho_5^{2/3}\dot M_{-10}\over
\Lambda_{ei}\avZ}\left({M_\odot\over
M}\right)\left({2\over\mu_e}\right)^{8/3},
\end{equation}
which depends only weakly on depth.

Equation (\ref{eq:rat1}) shows that $t_{\rm diff}\approx t_{\rm accr}$
for accretion rates of order $10^{-10}\ M_\odot\ {\rm yr^{-1}}$. This
is roughly the same critical accretion rate as found for the
degenerate ocean of an accreting neutron star in CZB (compare eq.~[29]
of that paper). The reason is that $t_{\rm diff}/t_{\rm accr}\propto
1/gR^2$, so that it depends only on the mass $M$ and not separately on
the gravity and radius.

At densities $\lesssim 10^3\ {\rm g\ cm^{-3}}$, the electrons become
non-degenerate, in which case the conductivity is given by Spitzer's
formula (Spitzer 1962)
\begin{equation}\label{eq:spitzer}
\sigma=7.5\times 10^{18}\ {\rm s^{-1}}\ {T_7^{3/2}\over \Lambda_{ei}\avZ},
\end{equation}
where the temperature $T=T_710^7\ {\rm K}$. For an ideal gas, the
pressure scale height is $H=P/\rho g=k_BT/\mu m_pg$, where $\mu$ is
the mean molecular weight, giving $H=8.3\times 10^6\ {\rm cm}\ T_7/\mu
g_8$. The ohmic time is
\begin{equation}
t_{\rm diff}=2.3\times 10^5\ {\rm yrs}\ {T_7^{7/2}\over \mu^2g_8^2\Lambda_{ei}\avZ}
\end{equation}
and the accretion time is
\begin{equation}
t_{\rm accr}=5.1\times 10^4\ {\rm yrs}\ {T_7\rho_2R_9^2\over \mu g_8\dot M_{-10}}.
\end{equation}
The ratio of the two is
\begin{equation}\label{eq:ratnd}
{t_{\rm diff}\over t_{\rm accr}}=3.3\ {T_7^{5/2}\over\rho_2}
{\dot M_{-10}\over\mu g_8\Lambda_{ei}\avZ}
\left({M_\odot\over M}\right).
\end{equation}
We have inserted typical values for temperature and density, although
the exact value of this ratio depends on the $T$--$\rho$ relation in
the atmosphere. For a constant flux atmosphere with free-free opacity,
$T\propto \rho^{4/13}$, giving $t_{\rm diff}/t_{\rm accr}\propto
\rho^{-3/13}$, again weakly dependent on density.

\begin{figure}
\begin{center}\epsfig{file=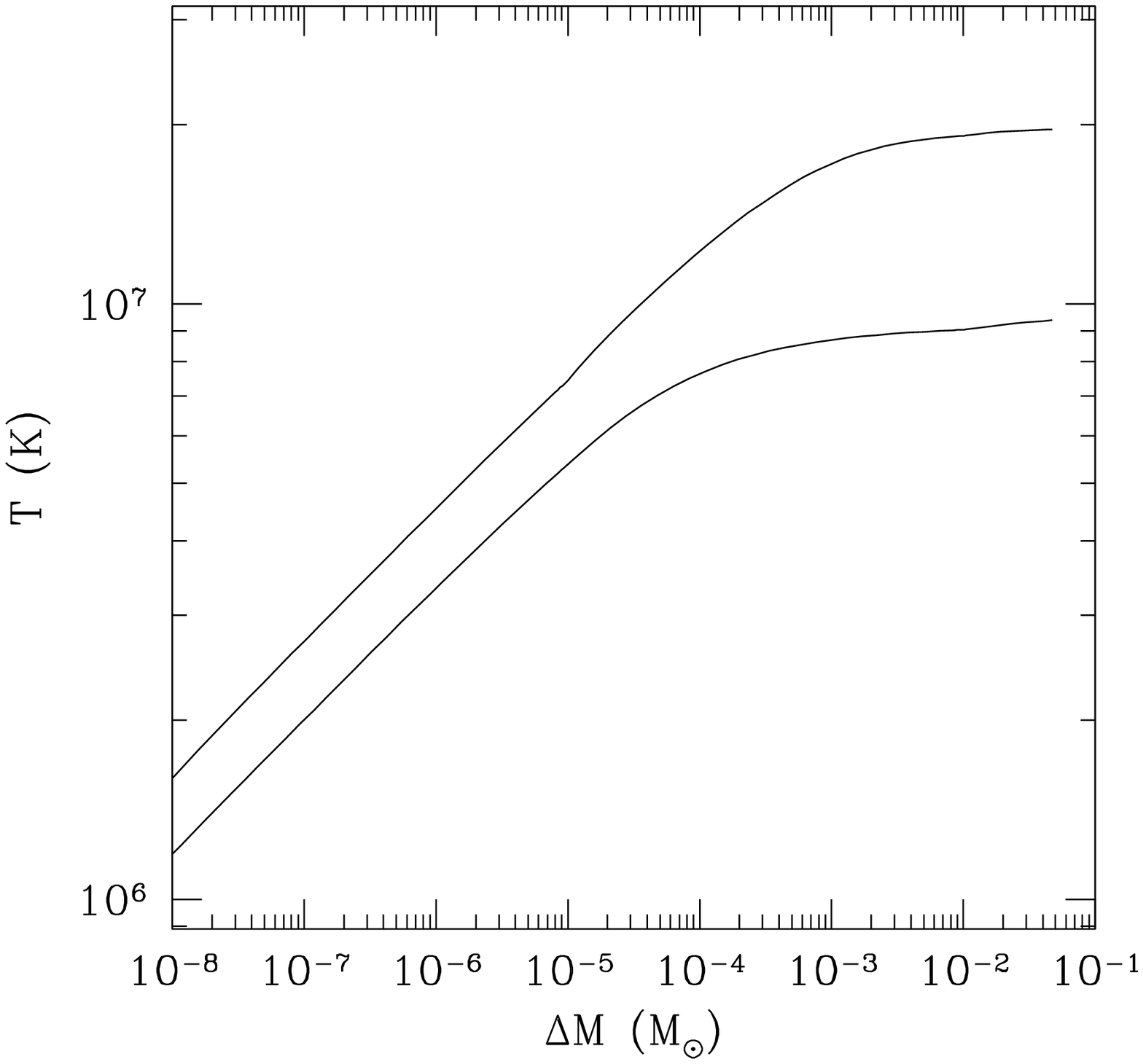,width=8.5cm}
\epsfig{file=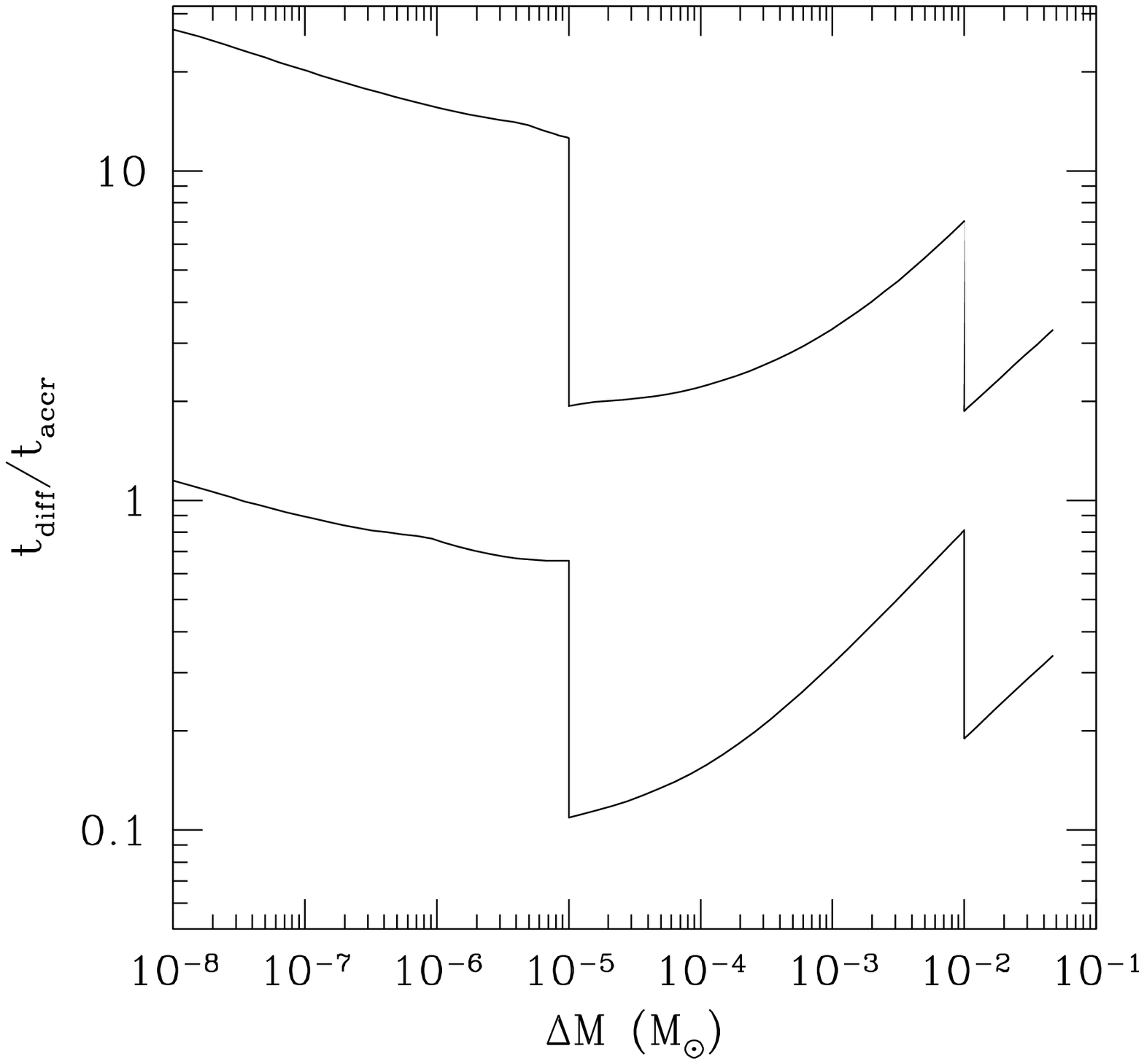, width=8.5cm}
\end{center}
\caption{ (i) Temperature profiles in the white dwarf envelope for
$\dot M=10^{-10}$ (lower curve) and $10^{-9}\ M_\odot\ {\rm yr^{-1}}$
(upper curve), for a $0.6 M_\odot$ white dwarf. (ii) Ratio of ohmic
time to accretion time in the envelope for $\dot M=10^{-10}$ (lower
curve) and $10^{-9}\ M_\odot\ {\rm yr^{-1}}$ (upper curve). The
composition is solar for $\Delta M<10^{-5}\ M_\odot$, pure He for
$10^{-5}\ M_\odot<\Delta M<10^{-2}\ M_\odot$, and C/O for $\Delta
M>10^{-2}\ M_\odot$.
\label{fig:env1}}
\end{figure}

\begin{figure}
\begin{center}
\epsfig{file=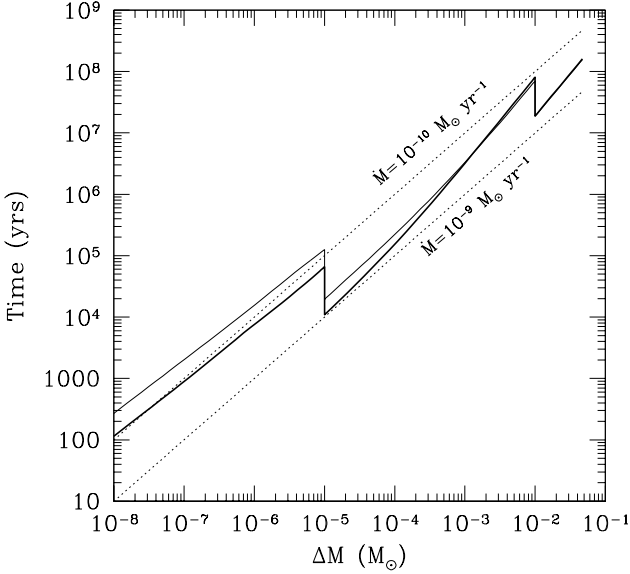, width=8.5cm}
\end{center}
\caption{ Ohmic diffusion time (solid lines) and accretion time
(dotted lines) in the white dwarf envelope for the models of Figure
\ref{fig:env1} ($\dot M=10^{-10}$ and $10^{-9}\ M_\odot\ {\rm
yr^{-1}}$). The accretion time (dotted lines) is labelled with the
appropriate accretion rate (the accretion time is smaller for a higher
accretion rate). The diffusion time (solid lines) is independent of
accretion rate in the degenerate layers (the conductivity depends only
on density); but is longer for higher accretion rates in the
non-degenerate layers (the conductivity increases with
temperature). \label{fig:env2}}
\end{figure}

\subsection{Detailed models}

We calculate models of the atmosphere by integrating the heat flux
equation
\begin{equation}\label{eq:heatflux}
F={4acT^3\over 3\kappa}{dT\over dy}
\end{equation}
and the entropy equation
\begin{equation}\label{eq:entropy}
{dF\over dy}=-{c_PT\dot m\over y}
\left[\nabla_{\rm ad}-{d\ln T\over d\ln y}\right],
\end{equation}
where $\kappa$ is the opacity, $c_P$ the specific heat at constant
pressure, and the adiabatic index is $\nabla_{\rm ad}=(d\ln T/d\ln
P)_{\rm ad}$. The terms on the right of equation (\ref{eq:entropy})
represent the effects of compressional heating (Nomoto 1982; Townsley
\& Bildsten 2001,2002).

We integrate equations (\ref{eq:heatflux}) and (\ref{eq:entropy}) from
the top of the atmosphere to the base, iterating the choice of flux at
the top until we match the correct flux from the core. We calculate
the opacity, which includes contributions from free-free, electron
scattering and conduction, as described by Schatz et al.~(1999). Our
calculation of electrical conductivity, which reduces to equation
(\ref{eq:sig}) for degenerate electrons, and equation
(\ref{eq:spitzer}) for non-degenerate electrons, is described in the
Appendix.

As an illustrative model of the outer layers of an accreting white
dwarf, we take a layer of solar abundance material of mass $10^{-5}\
M_\odot$, a pure helium layer of mass $10^{-2}\ M_\odot$, and a white
dwarf mass of $0.6\ M_\odot$, with an equal mixture by mass of carbon
and oxygen in the core. We choose a luminosity from the core of
$10^{-3}\ L_\odot$.  We show the resulting temperature profiles for
$\dot M=10^{-10}$ and $10^{-9}\ M_\odot\ {\rm yr^{-1}}$ in Figure
\ref{fig:env1}. For $\dot M=10^{-10}\ M_\odot\ {\rm yr^{-1}}$, we find
a core temperature $9.4\times 10^6\ {\rm K}$, luminosity $4.7\times
10^{-3}\ L_\odot$ and effective temperature $13,600\ {\rm K}$. For
$\dot M=10^{-9}\ M_\odot\ {\rm yr^{-1}}$, we find a core temperature
$2.0\times 10^7\ {\rm K}$, luminosity $6.2\times 10^{-2}\ L_\odot$ and
effective temperature $25,900\ {\rm K}$. These values are in
reasonable agreement with the detailed models of compressional heating
recently presented by Townsley \& Bildsten (2001, 2002). A simple estimate
of the compressional heating luminosity is to write
\begin{equation}
L\approx c_PT\dot M\approx 3.4\times 10^{-3}\ L_\odot\ T_7\dot M_{-10},
\end{equation}
where we use the heat capacity of an ideal gas $c_P=5k_B/2\mu m_p$,
and take $\mu\approx 1$ as a mean value in the envelope. This estimate
agrees well with our numerical results.

The ratio $t_{\rm diff}/t_{\rm accr}$ is shown in Figure
\ref{fig:env1}, and the ohmic time and accretion time individually in
Figure \ref{fig:env2}. Our numerical results agree well with the
analytic estimates in \S \ref{sec:analytic}. In the degenerate layers
($\Delta M\gtrsim 10^{-5}$), $t_{\rm diff}/t_{\rm accr}$ increases
with depth (eq.~[\ref{eq:rat1}]). The jumps in $t_{\rm diff}/t_{\rm
accr}$ at $\Delta M=10^{-5}\ M_\odot$ and $\Delta M=10^{-2}\ M_\odot$
result from the changes in composition from solar to pure He and from
pure He to C/O (eqs.~[\ref{eq:rat1}] and [\ref{eq:ratnd}] give $t_{\rm
diff}/t_{\rm accr}\propto 1/\avZ$). At low densities, $t_{\rm
diff}/t_{\rm accr}$ decreases with depth, as expected from
equation~(\ref{eq:ratnd}).

\begin{figure*}
\begin{center}
\epsfig{file=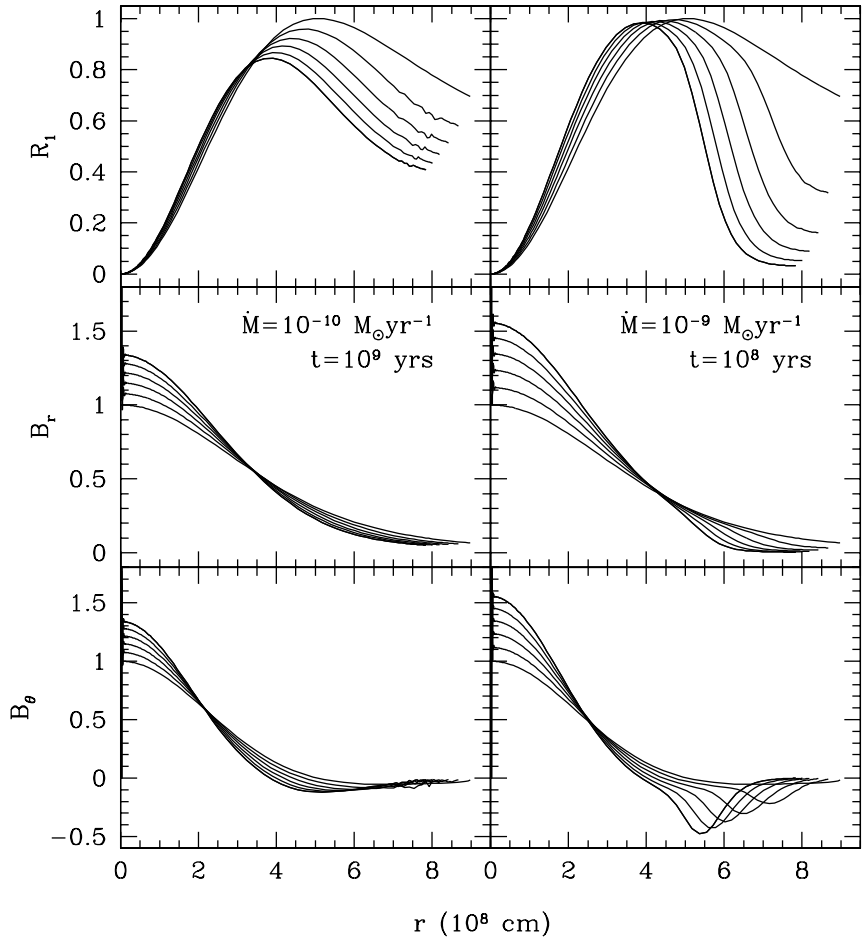, width=15 cm}
\end{center}
\caption{ Time evolution of the magnetic field during accretion of
$0.1\ M_\odot$ onto a $0.6\ M_\odot$ white dwarf. We assume ``vacuum''
boundary conditions (see discussion in \S 4.1). The left panel shows
$\dot M=10^{-10}\ M_\odot\ {\rm yr^{-1}}$, for which the accretion
takes $10^9$ years; the right panel shows $\dot M=10^{-9}\ M_\odot\
{\rm yr^{-1}}$, for which the time elapsed is $10^8$ years. The curves
are spaced by equal increments in central density, which increases
from $\approx 3\times 10^6\ {\rm g cm^{-3}}$ to $\approx 6\times 10^6\
{\rm g cm^{-3}}$. We normalise $R_1$ to have a maximum value $R_1=1$
initially, and normalise $B_r$ and $B_\theta$ to have a central value
of unity initially.
\label{fig:advectfig1}}
\end{figure*}

The condition $t_{\rm diff}=t_{\rm accr}$ defines a critical accretion
rate $\dot M_c$. Inspection of Figure \ref{fig:env1} shows that for
the accreted envelope ($\Delta M<10^{-5}\ M_\odot$), $\dot M_c\approx
10^{-10} M_\odot\ {\rm yr^{-1}}$, and for the He layer and the C/O
core, an average value is $\dot M_c\approx 5\times 10^{-10} M_\odot\
{\rm yr^{-1}}$. This agrees well with our analytic estimates (compare
eqs.~[\ref{eq:rat1}] and [\ref{eq:ratnd}] when $t_{\rm diff}/t_{\rm
accr}=1$).

\section{Global Models of Accretion and Diffusion}

In this section, we follow the effects of advection and diffusion in a
simple model of the global magnetic field structure. We show that the
evolution is very different depending on whether $\dot M$ is less than
or greater than the critical rate $\dot M_c\approx (1$--$5)\times
10^{-10}\ M_\odot\ {\rm yr^{-1}}$ found in \S 3.

\subsection{A simple model of the global field evolution}

We make the following simplifying assumptions: (i) we consider an
axisymmetric, poloidal magnetic field, (ii) we assume spherical
accretion, and (iii) we take the accreted matter to have the same
composition as the core (equal amounts by mass of C and O). Under
spherical accretion, each spherical harmonic component $l$ evolves
independently, so that an initially dipolar field remains dipolar as
accretion proceeds. In reality, the accreted matter is channelled onto
the polar caps in magnetic systems, subsequently spreading over the
surface of the star. We do not attempt to model this process here,
meaning that we must pay careful attention to our choice of surface
boundary condition. We discuss this issue below, but first outline the
equation describing the evolution of the magnetic field, and how we
solve it.

The induction equation with advection included is
\begin{equation}\label{eq:induction}
{\partial\vec{B}\over\partial
t}=\curl\left(\vec{v}\vcross\vec{B}\right)-\curl\left(\eta\curl\vec{B}\right),
\end{equation}
or, in terms of the vector potential $\vec{A}$,
\begin{equation}\label{eq:Av}
{\partial\vec{A}\over\partial
t}=\vec{v}\vcross\left(\curl\vec{A}\right)-\eta\curl\curl\vec{A}.
\end{equation}
As previously, we consider an axisymmetric field, so that
$\vec{A}=A_\phi(r,\theta)\hat{e}_\phi$. In the absence of diffusion,
equation (\ref{eq:Av}) gives
\begin{equation}\label{eq:advect}
\left({\partial\over\partial t}+\vec{v}\vdot\grad\right)
\left(rA_\phi\sin\theta\right)=0
\end{equation}
(WVS; Choudhuri \& Konar 2002). The quantity $rA_\phi\sin\theta$
(which labels the magnetic field lines, see \S 2.2) is advected by the
poloidal velocity field.

As before, we separate $A_\phi$ into radial and angular pieces
according to equation (\ref{eq:expand}). We take the velocity field to
be purely radial, $\vec{v}=v_r(r,\theta)\hat{e}_r$. Equation
(\ref{eq:Av}) then gives
\begin{equation}\label{eq:dRdt}
{\partial R_l\over\partial t}=-v_r{\partial R_l\over \partial r}+
\eta\left[{\partial^2R_l\over\partial r^2}-{l(l+1)R_l\over
r^2}\right].
\end{equation}
This equation governs the evolution of $R_l(r,t)$ under the joint
action of accretion and diffusion.

We evolve equation (\ref{eq:dRdt}) numerically\footnote{This numerical
approach is similar to that of WVS, who included the advection terms
in their study of cooling white dwarfs. WVS found that the significant
contraction which occurs in the pre-white dwarf stages of evolution
increased the rate of ohmic decay, by taking the initial $n=1$ decay
mode and generating higher $n$ components. This interesting effect is
not significant during the accretion process, which involves much
smaller changes in white dwarf radius.}. At each time step, we include
advection by making a new white dwarf model (as described in \S 2.1)
with a slightly larger mass, but keep $R_l$ fixed for a given fluid
element. We then apply an implicit Crank-Nicholson scheme for the
diffusion term (e.g.~Press et al.~1992). With accretion switched off,
our code accurately follows the predicted decay of the eigenmodes
found in \S 2.

We adopt two different boundary conditions at the surface. The first
is the ``vacuum'' boundary condition given by equation (\ref{eq:bc2}).
The second boundary condition is a ``screened'' boundary condition
$R_l=0$ at the surface. In the advection step, the boundary condition
determines the values of $R_l$ assigned to the newly accreted
matter. As described above, we take the composition of the accreted
material to be equal mass fractions of C and O. At the centre of the
star, where $v_r$ vanishes, we apply equation (\ref{eq:bc1}).

The ``vacuum'' boundary condition implies $\vec{J}=0$ outside the
star, as in \S 2, and that each shell of matter is accreted with the
vacuum field at the surface. This boundary condition was adopted by
Choudhuri \& Konar (2002) in their recent study of accreting neutron
stars (these authors solved eq.~[\ref{eq:Av}] with a prescribed 2D
velocity field). In our 1D model, we envisage this boundary condition
as approximating the case where the accreted material is able to
spread away from the polar cap without significantly distorting the
field structure. This is similar to Choudhuri \& Konar (2002), who put
in an outwards radial velocity in the outer layers to simulate the
effects of buoyancy instabilities. They found the evolution of the
field was then determined by advection in the interior. The
``screened'' boundary condition $R_l=0$ assumes screening currents at
the surface, similar in spirit to the plane parallel models of CZB for
accreting neutron stars, and to the idea that the magnetic field may
be completely buried by accretion.

These two different boundary conditions illustrate the range of
behaviour that might be expected given a more complex flow geometry
and better treatment of the outer layers, including effects such as
buoyancy or interchange instabilities. However, note that the solution
for the interior is insensitive to the choice of surface boundary
condition. In the next section, we present our numerical results. We
show that the evolution of the field depends upon whether the
accretion rate is above or below the critical rate $\dot M_c$.

\begin{figure*}
\begin{center}
\epsfig{file=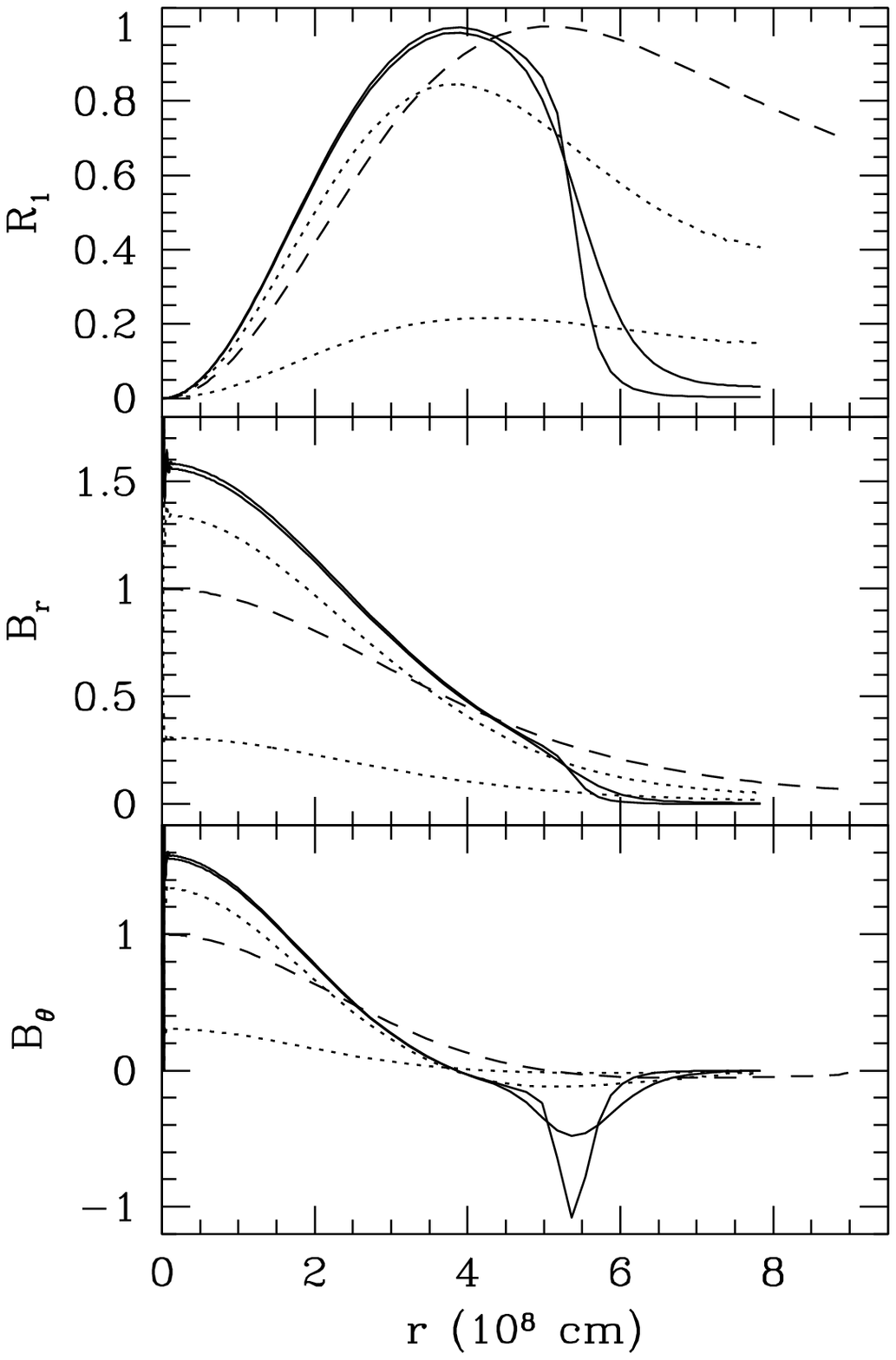, width=7.5cm}
\epsfig{file=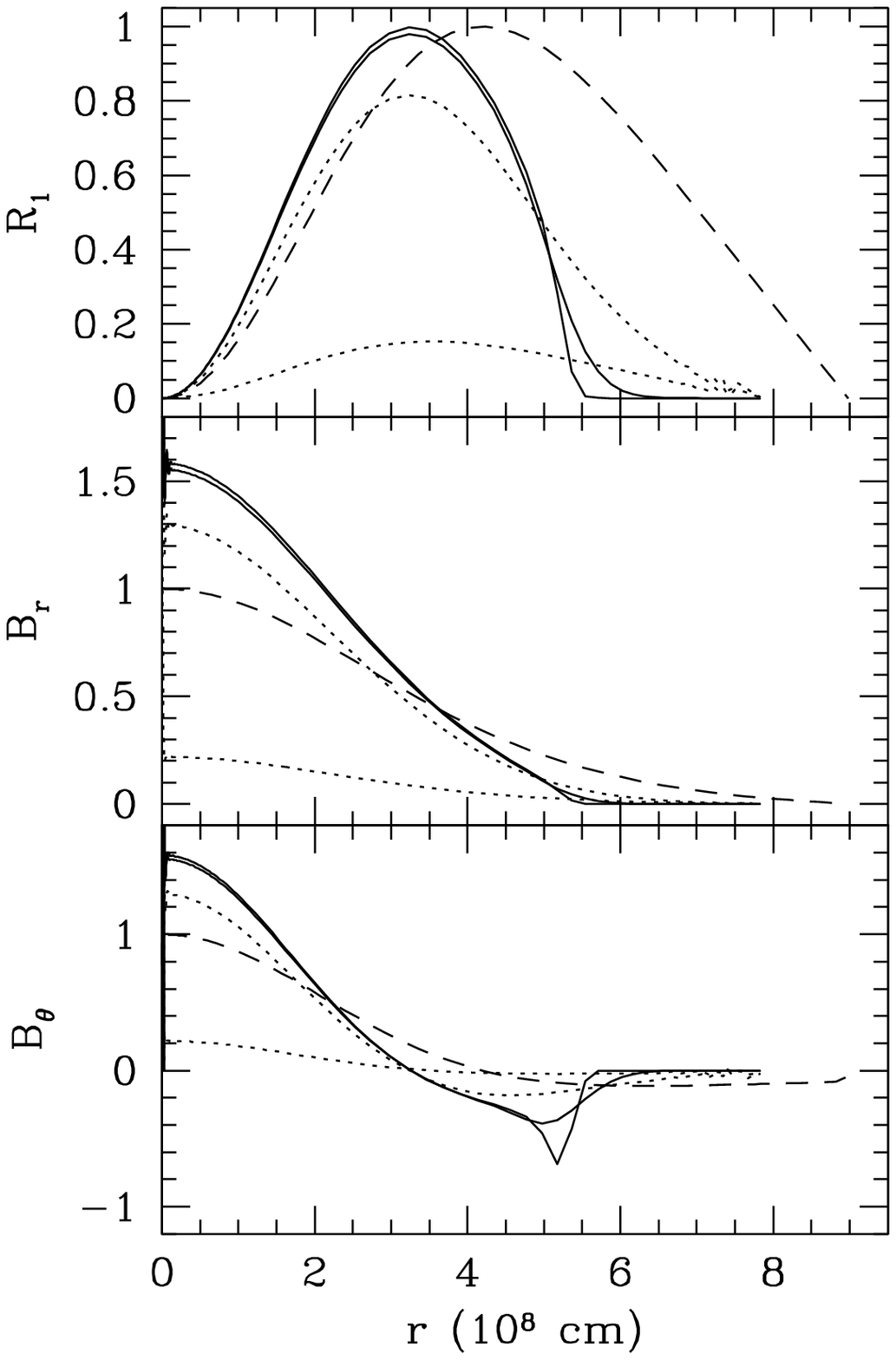, width=7.5cm}
\end{center}
\caption{ The profiles of $R_1$, $B_r$, and $B_\theta$ after the
accretion of $0.1\ M_\odot$ onto a $0.6\ M_\odot$ white dwarf at
different rates. We show results for vacuum boundary conditions (left
panel) and screening boundary conditions (right panel). The dashed
line in each panel shows the initial state. The solid lines show
accretion rates $10^{-8}$ and $10^{-9}\ M_\odot\ {\rm yr^{-1}}$, for
which mass is accreted too rapidly to become significantly
magnetised. The dotted lines show accretion rates $10^{-10}$ and
$10^{-11}\ M_\odot\ {\rm yr^{-1}}$, for which ohmic diffusion has time
to magnetise the accreted material. Significant ohmic decay of the
field occurs at these low rates, because the accretion takes a long
time.
\label{fig:advectend}}
\end{figure*}

\begin{figure}
\begin{center}
\epsfig{file=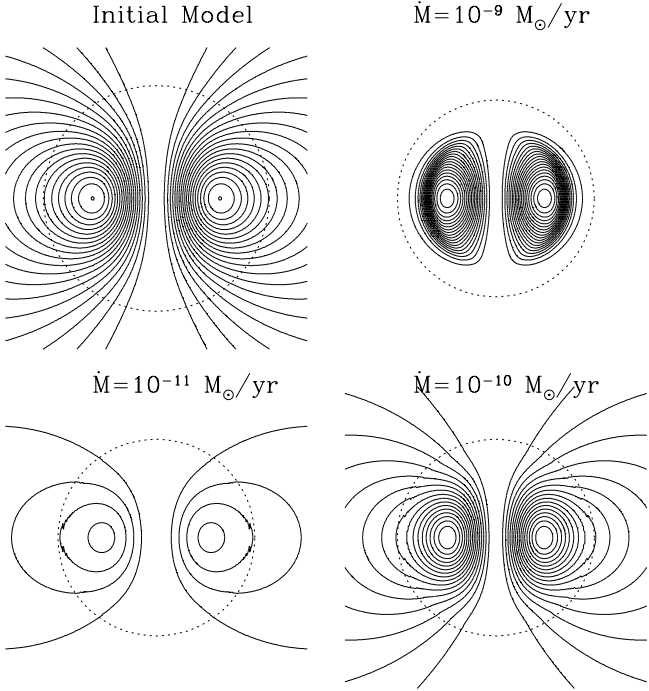, width=9cm}
\end{center}
\caption{Magnetic field lines for the models shown in the left panel
of Figure 6. We show the initial field structure in the top left
panel, and the final structure for (clockwise from the top right)
$\dot M=10^{-9}$, $10^{-10}$, and $10^{-11}\ M_\odot\ {\rm
yr^{-1}}$. The dotted line indicates the stellar surface (note the
larger radius of the less massive initial model). We show field lines
with the same value of flux in each panel. The dramatic ohmic decay at
the lowest accretion rate appears as the reduction in the number of
field lines.
\label{fig:cont}}
\end{figure}

\begin{figure}
\begin{center}
\epsfig{file=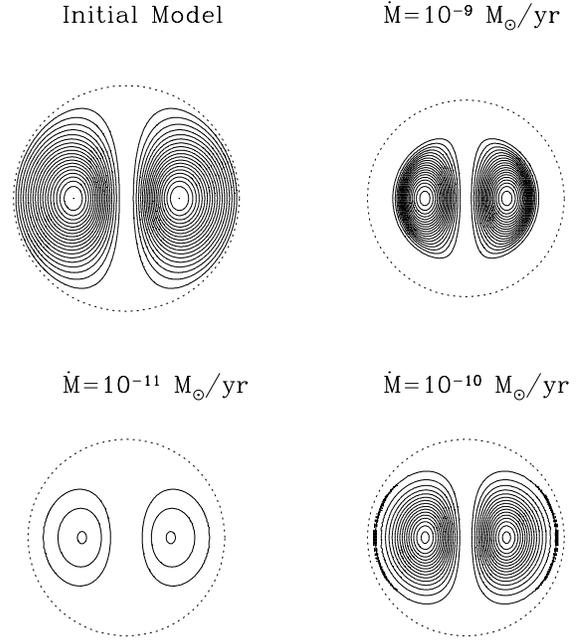, width=9cm}
\end{center}
\caption{As Figure \ref{fig:cont}, but for the right panel of Figure 6
(screening boundary condition).
\label{fig:cont2}}
\end{figure}

\subsection{Results}

Figure \ref{fig:advectfig1} shows the time evolution of a dipole field
during accretion of $0.1\ M_\odot$ onto a $0.6 M_\odot$ white dwarf at
two different accretion rates. We take a ``vacuum'' boundary
condition, and the initial profile of $R_1(r)$ to be the lowest order
dipole decay mode. We show the time evolution of $R_1$, $B_r$
($\propto R_1/r^2$, eq.~[\ref{eq:Br}]), and $B_{\theta}$ ($\propto
(1/r)(dR_1/dr)$, eq.~[\ref{eq:Bt}]) at equal steps in central
density. We normalize $R_1$ so that the initial profile has a maximum
value $R_1=1$, and normalize $B_r$ and $B_\theta$ so that the initial
central value is unity. The initial central density is $\approx
3\times 10^6\ {\rm g\ cm^{-3}}$, increasing to $\approx 6\times 10^6\
{\rm g\ cm^{-3}}$. The left panel is for $\dot M=10^{-10}\ M_\odot\
{\rm yr^{-1}}$, in which case the accretion lasts for $10^9$ years,
and the right panel is for $\dot M=10^{-9}\ M_\odot\ {\rm yr^{-1}}$,
in which case the accretion lasts for $10^8$ years.

At $\dot M=10^{-10}\ M_\odot\ {\rm yr^{-1}}$ (left panel of Figure
\ref{fig:advectfig1}), the magnetic field has time to diffuse into the
newly accreted material. The overall decrease in $R_1$ is due to ohmic
decay during the $10^9$ years of accretion (the lowest order decay
time of the initial model is $\tau_{11}=7.6$ billion years). At $\dot
M=10^{-9}\ M_\odot\ {\rm yr^{-1}}$ (right panel of Figure
\ref{fig:advectfig1}), the accreted material is added too quickly for
significant diffusion, so that the surface value of $R_1$ drops as
accretion proceeds. The central values of $B_r$ and $B_\theta$
increase by a factor of $\approx 1.6$. This can by understood in terms
of the combination of flux conservation, which gives $r^2B=$constant,
and mass conservation, which gives $r^3\rho=$constant. Eliminating
$r$, we find $B\propto\rho^{2/3}$, giving an increase in $B$ of 1.6
when $\rho$ increases by a factor of 2. (In the thin layer considered
by CZB, the compression was one-dimensional, giving $B\propto\rho$ in
that case; see their eq.~[13]).

Figure \ref{fig:advectend} shows the final magnetic profiles for
accretion at several different rates. We show the profiles for vacuum
boundary conditions (left panel) and screened boundary conditions
(right panel). The initial profile in each case is shown by the dotted
line, and is the lowest order decay mode of the initial model (subject
to the respective surface boundary condition). We show accretion rates
$\dot M=10^{-11}$, $10^{-10}$, $10^{-9}$ and $10^{-8}\ M_\odot\ {\rm
yr^{-1}}$. At the lowest accretion rates, $10^{-11}$ and $10^{-10}\
M_\odot\ {\rm yr^{-1}}$ (dashed lines), the magnetic field is able to
diffuse into the newly accreted layers. The decrease in $R_1$ is due
to ohmic decay over the long timescale of accretion. For accretion
rates $10^{-9}$ and $10^{-8}\ M_\odot\ {\rm yr^{-1}}$ (solid lines),
the accreted layers do not have time to become significantly
magnetized. At $\dot M=10^{-8}\ M_\odot\ {\rm yr^{-1}}$, the final
profile is almost exactly that given by conserving $R_1$ in each fluid
element as accretion proceeds.

In Figures \ref{fig:cont} and \ref{fig:cont2}, we show the magnetic
field lines for the profiles in Figure \ref{fig:advectend}. We show
field lines with the same values of $rA_\phi\sin\theta$ in each
panel. The $\dot M=10^{-11}\ M_\odot\ {\rm yr^{-1}}$ has only two
field lines because of the significant ohmic decay that occurs as
accretion proceeds. Similarly, at $10^{-9}\ M_\odot\ {\rm yr^{-1}}$,
all of the field lines shown in the initial model have been pushed
into the interior by the accretion flow.


\begin{figure}
\begin{center}
\epsfig{file=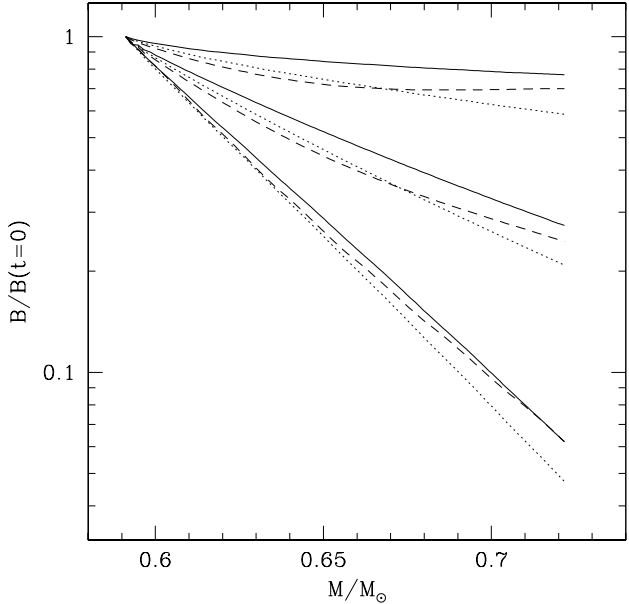, width=8.5cm}
\end{center}
\caption{ For the vacuum boundary condition solutions, we show surface
values of $R_1$ (dotted line), $B_r$ (solid line), and $B_\theta$
(dashed line) relative to their initial values as a function white
dwarf mass for (top to bottom) $\dot M=10^{-10}$, $10^{-11}$, and
$10^{-9}\ M_\odot\ {\rm yr^{-1}}$.
\label{fig:advectsurf}}
\end{figure}

Figure \ref{fig:advectsurf} shows $R_1$, $B_r$, and $B_\theta$ at the
surface as a function of mass accreted for a vacuum boundary
condition. As we discussed \S 4.1, the quantitative decrease in the
surface field depends on the chosen surface boundary
condition. However, Figure \ref{fig:advectsurf} shows the qualitative
effect of accretion at different rates. From top to bottom, the
accretion rates are $\dot M=10^{-10}$, $10^{-11}$, and $10^{-9}\
M_\odot\ {\rm yr^{-1}}$. At $\dot M=10^{-10}\ M_\odot\ {\rm yr^{-1}}$,
the surface field changes only slightly as the newly accreted material
is able to become magnetized by diffusion. For $\dot M=10^{-11}\
M_\odot\ {\rm yr^{-1}}$, the surface field decreases due to ohmic
decay (in this case the accretion takes $10^{10}$ years, comparable to
the decay time). For $\dot M=10^{-9}\ M_\odot\ {\rm yr^{-1}}$, the
change is directly due to accretion.

\section{Implications for Observed Systems}

In the previous sections, we have shown that, under the assumption
that the white dwarf retains the mass it accretes, the surface
magnetic field may be significantly reduced for accretion rates
greater than the critical value $\dot M_c\approx (1$--$5)\times
10^{-10}\ M_\odot\ {\rm yr^{-1}}$. We now conclude by discussing the
implications of our results for observed systems. We show that there
are several interesting consequences of abandoning the assumption that
the magnetic fields of accreting white dwarfs do not change with time.

\begin{figure}
\begin{center}
\epsfig{file=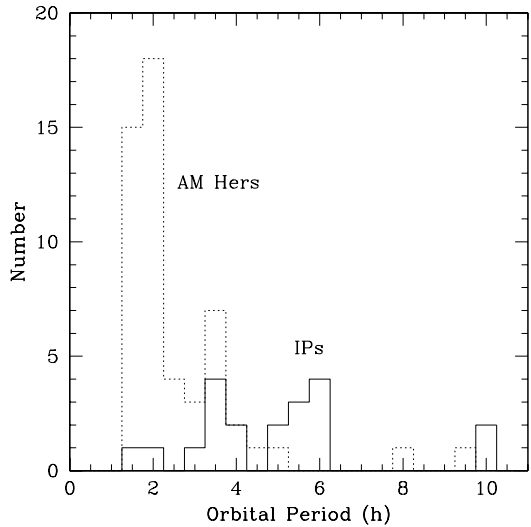, width=8.5cm}
\end{center}
\caption{
The orbital periods of AM Her systems and intermediate polars. We show
orbital periods for 53 AM Her systems from Wickramasinghe \& Ferrario
(2000), and for 20 intermediate polars from the catalogue of Ritter \&
Kolb (1998). We exclude the intermediate polar GK Per which has
$P_{\rm orb}=2\ {\rm days}$ and an evolved companion.
\label{fig:periodhist}}
\end{figure}

\subsection{An evolutionary connection between AM~Hers and intermediate polars?}

Accreting white dwarfs fall into two classes. Those in AM Her systems
rotate synchronously with the binary orbital period, and have fields
$10^7$--$3\times 10^8\ {\rm G}$ (WF). Intermediate polars (see review
by Patterson 1994), in which the white dwarf rotates asynchronously,
are believed to contain white dwarfs with fields $\sim 10^5$--$10^7\
{\rm G}$, although most have not been measured directly (the lower
limit here is the field strength needed to disrupt the accretion flow
before it hits the white dwarf surface). The asynchronous rotation of
the white dwarfs in IPs lead to early suggestions that the white dwarf
magnetic fields are lower than in AM Hers (for example, Lamb \&
Patterson 1983 estimated $B\sim 10^6\ {\rm G}$). In addition, whereas
strong optical polarization is seen in AM Hers, little or no
polarization is seen in IPs, also suggesting weaker magnetic
fields. Recently, the lack of Zeeman splitting in spectroscopic
observations of the white dwarf in V709~Cas (Bonnet-Bidaud et
al.~2001) has constrained the magnetic field to be $<10^7\ {\rm G}$.

The two classes show an interesting difference in accretion rates. AM
Her systems accrete at low rates, $\dot M\approx 5\times 10^{-11}\
M_\odot {\rm yr^{-1}}$ (Chanmugam, Ray, \& Singh 1991; Warner 1995),
whereas the IPs accrete rapidly, Warner (1995) estimates $\dot
M\approx (0.2$--$4)\times 10^{-9}\ M_\odot\ {\rm yr^{-1}}$ from X-ray
fluxes of ten systems. Wickramasinghe \& Wu (1994) propose that the
low accretion rate of AM Hers is a direct consequence of the strong
magnetic field of the white dwarf, which inhibits magnetic braking by
reducing the number of open field lines along which the magnetized
wind from the secondary may flow (for example, see Li, Wickramasinghe,
\& Wu 1995).

Figure \ref{fig:periodhist} shows the orbital period distributions of
known AM Hers and IPs. Most AM Her systems have $P_{\rm orb}\lesssim
3$ hours, whereas most IPs have $P_{\rm orb}\approx 3$--$6$
hours. This difference has led many to suggest an evolutionary link
between the two classes, namely that the asynchronous IPs are the
progenitors of the synchronous AM Her systems. For a magnetic field
$B\sim 10^7\ {\rm G}$, the magnetospheric radius equals the orbital
separation at $P_{\rm orb}\approx 4\ {\rm hrs}$, suggesting a natural
dividing line between synchronous and asynchronous systems (Chanmugam
\& Ray 1984; King, Frank, \& Ritter 1985; Hameury et
al.~1987). However, an evolutionary connection is difficult if the
white dwarf magnetic field is constant with time (e.g., Wu \&
Wickramasinghe 1993). It requires either (i) the IPs and AM Hers have
similar magnetic field strengths, but the optical polarization is
somehow suppressed in asynchronous rotators, or (ii) the IPs have
weaker fields than the AM Her systems, implying a yet to be detected
population of asynchronous AM Her progenitors (King \& Lasota 1991).
 
It is striking that the mean accretion rate of the IPs is $>\dot M_c$,
whereas the AM Her systems accrete at rates $<\dot M_c$. Thus, given
our results, a natural suggestion is that the magnetic fields of the
intermediate polars are low because of the direct effects of the
accretion flow. If so, it is possible for an intermediate polar to
evolve into an AM Her system if the mass transfer rate to drops from
$>\dot M_c$ to $<\dot M_c$ at an orbital period $\approx (3$--$4)$
hours (between the observed AM Her and intermediate polar
populations).

The reason for such a change in accretion rate is not clear, but is
reminiscent of the need for a rapid drop in mass transfer rate to
explain the period gap ($P_{\rm orb}\approx 2$--$3$ hours) in
non-magnetic CVs. Whether or not the magnetic CVs exhibit a period gap
is unclear, there are certainly a higher fraction of magnetic systems
in the gap than for non-magnetic systems (King 1994; Kolb 1995;
Wheatley 1995). It is however possible that a similar mechanism is
operating in both cases, perhaps disrupted magnetic braking, as
preferred for non-magnetic systems (Rappaport, Verbunt \& Joss 1983;
Howell, Nelson, \& Rappaport 2001).

There is an alternative scenario along the same lines as that of
Wickramasinghe \& Wu (1994). The ROSAT detection of IPs with soft
X-ray spectra similar to weaker field AM Hers (Haberl \& Motch 1995)
has led to suggestions of two populations of IPs. A possibility is
that those IPs with fields close to $10^7$ G come into synchronous
rotation at $P_{\rm orb}\approx 4$ hours, which causes a drop in $\dot
M$ as suggested by Wickramasinghe \& Wu (1994), allowing the field to
subsequently grow to values $\sim 10^8\ {\rm G}$ typical of AM Hers. A
problem with this picture is what happens to the weaker field IPs
(i.e., those with traditional hard X-ray spectra), since very few of
them are seen at short orbital periods.
 
The timescale for reemergence of the field is set by the ohmic
diffusion time at the base of the accreted layer. Figure 4 and
equation (16) show that this is $\approx 3\times 10^8\ {\rm years}\
(\Delta M/0.1 M_\odot)^{7/5}$ where $\Delta M$ is the accreted
mass. This is similar to the time for a non-magnetic CV to cross the
period gap (Howell et al.~2001). Thus if the drop in mass accretion
rate causes the secondary to lose contact with the Roche lobe, there
may be time for the field to reemerge before contact is resumed. The
impact of any gap on the observed orbital period distribution depends
on how strongly the transition orbital period depends on mass transfer
rate and magnetic field. With or without a period gap, the lack of
known synchronous low field systems could constrain the timescale for
reemergence and therefore accreted mass\footnote{We thank the referee,
J.-M. Hameury, for bringing up this issue.}.

A detailed test of either scenario requires comparison of evolutionary
studies and the observed populations of IPs and AM Hers. We note that
whereas there is a clear difference in orbital periods and magnetic
field strengths between AM Hers and IPs as classes of objects, no
correlation is observed between magnetic field strength and orbital
period amongst AM Her systems only (WF).

\subsection{Where are the $10^9\ {\rm G}$ accreting WDs?}

The highest measured magnetic field in an accreting system is
$2.3\times 10^8\ {\rm G}$ for AR Uma, whereas the highest observed
fields in isolated WDs are close to $10^9\ {\rm G}$. Whether or not
this difference is significant remains to be seen. For example, WF
argue that selection effects hide the very high field accreting
WDs. There have been suggestions as to why the observed upper limit
might differ. Wickramasinghe \& Wu (1994) argue that because the high
field systems come into synchronization at wider separation, magnetic
braking is suppressed earlier, and they do not have time to evolve via
gravitational radiation into contact (thus they predict a population
of high field, detached binaries).

We would like to point out that some difference would be expected
simply because of the different ohmic decay properties of the two
populations. The liquid accreting white dwarfs decay at a faster rate
than the isolated white dwarfs, which form a solid core after $\sim
10^9$ years of cooling, effectively switching off ohmic decay. To find
the maximum difference between the two populations, we assume the
isolated white dwarfs undergo no magnetic field decay. The contrast in
field strengths is then $\exp(t/\tau)$, where $\tau$ is the decay time
for the accreting population. For $\tau=10^{10}\ {\rm years}$, this
gives a factor of 2 after $t=7$ billion years, or a factor of 3 after
$t=11$ billion years. This effect alone may not explain the observed
factor of $\approx 3$, but should be taken into account when comparing
the maximum field strengths of the two populations.

\subsection{Type Ia progenitor systems}

Currently favoured models for Type Ia supernovae progenitors are
rapidly accreting ($\dot M\gtrsim 10^{-7}$ $M_\odot\ {\rm yr^{-1}}$)
massive white dwarfs, which are able to burn the accreted hydrogen and
helium steadily and grow to near-Chandrasekhar mass. Examples are the
white dwarfs in supersoft X-ray sources and symbiotic binaries. If
accretion is acting to suppress the surface magnetic field in these
sources as we propose for intermediate polars, we would not expect to
find strongly magnetic white dwarfs in these systems. There is little
information about the incidence of magnetism in these rapidly
accreting systems. In a survey of 35 symbiotics, Sokoloski \& Bildsten
(1999) (see also Sokoloski, Bildsten, \& Ho 2001) discovered a
magnetic WD in the symbiotic binary Z And rotating with a 28 minute
spin period. If the white dwarf is in spin equilibrium with an
accretion disk at the magnetospheric radius, the implied magnetic
field strength is $6\times 10^6\ {\rm G}$, typical of an intermediate
polar. XMM observations of M31 revealed a supersoft X-ray source with
865s pulsations, interpreted as the spin of a magnetized white dwarf
(Osborne et al.~2001). A similar estimate for the magnetic field gives
$\approx 10^7\ {\rm G}$ (King, Osborne, \& Schenker 2002). It remains
to be seen how common systems like these are amongst rapidly accreting
sources.

As the central density increases due to accretion, the internal field
is amplified by compression ($B\propto\rho^{2/3}$; see \S 4.2). This
amplification is shown in Figure 5 for accretion onto a $0.6\ M_\odot$
white dwarf, and would be even greater for a massive white dwarf. The
density increases especially quickly as you approach Chandrasekhar
mass and the electrons become more relativistic, as pointed out by
work on compressional heating (Nomoto 1982; Townsley \& Bildsten 2001,
2002). An important problem is to understand the role played by such
an amplified magnetic field in the approach to and evolution after
ignition of a Type Ia supernova. For example, Ghezzi et al.~(2001),
recently pointed out that a large scale magnetic field of
$10^8$--$10^9\ {\rm G}$ can generate asymmetries in the thermonuclear
flame front that engulfs the white dwarf during the supernova. Our
results show that the prior evolution of the field during accretion
should be taken into account.

\subsection{Complexity as a result of accretion}

Both isolated and accreting white dwarfs often show a complex magnetic
field geometry, requiring a significant quadrupole component or an
offset dipole (WF). In \S 2, we found decay times of $(4$--$6)\times
10^9\ {\rm years}$ for the $l=2$ decay mode of liquid white
dwarfs. Thus one possibility is that the quadrupole field is fossil,
requiring that the dipole and quadrupole fields had comparable
magnitudes initially.

In isolated systems, the liquid decay time is relevant only for the
first $\sim 10^9\ {\rm yrs}$, after which the formation of a solid
core increases the ohmic time. Muslimov et al.~(1995) studied the Hall
effect in isolated white dwarfs as a possible mechanism for generating
field complexity as the white dwarf cools. However, the Hall effect is
not expected to play a significant role for a liquid interior
(Muslimov et al.~1995).

Another possibility is that the accretion flow directly leads to
complexity of the field. The spreading of material away from the polar
cap (which we have not included in our models in this paper) could
induce higher order components of the surface field. An interesting
observation is that the magnetic pole undergoing most accretion in AM
Her systems is the pole with the weaker magnetic field in all cases
with magnetic field measurements for both poles (WF). Wickramasinghe
\& Wu (1991) suggest that this is a result of the role of the
quadrupole component during synchronisation. An alternative suggested
by our results is that after synchronisation occurs, the pole pointing
towards the companion undergoes more rapid accretion which reduces the
field strength at that pole, giving rise to the observed
asymmetry. Even though the global rate for AM Hers is $<\dot M_c$, the
local accretion rate in the layers for which the accreted material is
still confined to the polar cap will be greater than the critical
rate. Detailed models of the spreading of material away from the polar
cap are required to investigate this further.

\section{Conclusions}

We have studied the evolution of the magnetic field in an accreting
white dwarf. Previous studies of ohmic decay in isolated white dwarfs
showed that the ohmic decay time is always longer than the cooling
time because the white dwarf develops a solid core. In \S 2, we
calculated the ohmic decay times for the lowest order modes of
accreting white dwarfs, which have a liquid interior because of
compressional heating by accretion. We found that the lowest order
ohmic decay time is $(8$--$12)\times 10^9\ {\rm yrs}$ for a dipole
field, and $(4$--$6)\times 10^9\ {\rm yrs}$ for a quadrupole field
(see Table 1), in good agreement with the earlier calculations of
WVS. The difference in ohmic decay times between isolated and
accreting white dwarfs should be taken into account when comparing the
maximum field strengths of the two populations. In addition, the decay
timescale for the quadrupole is long enough that observed quadrupole
components (see WF for a summary of the observations) in both
accreting and isolated white dwarfs may be fossil if the quadrupole is
initially of similar strength to the dipole component.

In \S 3, we compared the timescale for ohmic decay with accretion, and
showed that accretion occurs more rapidly than ohmic diffusion for
accretion rates greater than the critical rate $\dot M_c\approx
(1$--$5)\times 10^{-10}\ M_\odot\ {\rm yr^{-1}}$. In \S 4, we
calculated the time evolution of the magnetic field as a function of
accretion rate, assuming the white dwarf mass increases with time. For
a simplified field and accretion geometry (an axisymmetric poloidal
magnetic field and spherical accretion), we found that accretion at
rates $\dot M>\dot M_c$ leads to a reduction in the field strength at
the surface, as the field is advected into the interior by the
accretion flow.

The main conclusion of this paper is that, due to the direct action of
accretion, significant changes in the surface magnetic field of an
accreting white dwarf could occur during its accretion lifetime. In \S
5, we showed that accretion induced magnetic field evolution may
explain several features of observed systems. Most striking is that
the strongly magnetic ($B\sim 10^7$--$3\times 10^8\ {\rm G}$) AM Her
systems have a mean accretion rate $\approx 5\times 10^{-11}\ M_\odot\
{\rm yr^{-1}}<\dot M_c$ whereas the weakly magnetic ($B\sim
10^5$--$10^7\ {\rm G}$) IPs accrete at rates $\sim 10^{-9}\ M_\odot\
{\rm yr^{-1}}>\dot M_c$. This raises the possibility that the white
dwarfs in IPs have subsurface fields as strong as those in AM Hers. If
so, this allows for an evolutionary connection between the long
orbital period ($P_{\rm orb}\approx 3$--$6$ hours) IPs and short
orbital period AM Hers ($P_{\rm orb}\lesssim 3$ hours), requiring that
evolution in orbital period somehow causes a drop in accretion rate
(for example, by disrupted magnetic braking as postulated for
non-magnetic CVs; see \S 5.1 for further discussion). The magnetic
field would then reemerge on a timescale set by ohmic diffusion,
$\approx 3\times 10^8\ {\rm yrs}\ (\Delta M/0.1\ M_\odot)^{7/5}$, where
$\Delta M$ is the accreted mass. Evolutionary calculations are
necessary to compare this scenario with the properties of observed
systems. We did not consider non-magnetic accreting systems in \S 5;
however, it is possible that many non-magnetic white dwarfs (meaning
$B\lesssim 10^5\ {\rm G}$ so that the accretion flow is not disrupted)
have submerged magnetic fields if they are accreting at rates $>\dot
M_c$.

The models presented in this paper are simplified, and many
outstanding theoretical questions remain. The major uncertainty in the
evolution of accreting white dwarfs is whether the white dwarf mass
increases or decreases with time. Even if the mass is increasing with
time as we have assumed here, classical nova explosions eject some
mass from the system; the effective mass accretion rate is therefore
less than the mass transfer rate from the secondary by some factor,
introducing an extra uncertainty. If classical novae eject more mass
than accreted prior to the thermal runaway, excavation of the white
dwarf core leads to a decreasing mass with time. This raises the
possibility that the surface magnetic field could grow with time if
underlying magnetic field is exposed as the mass decreases. This was
originally suggested as a source for the magnetic field of DQ Her by
Lamb (1974). Interestingly, this alternative is also consistent with
an evolutionary link between AM Hers and intermediate polars, since
the magnetic field would increase as the orbital period decreases and
successive nova explosions occur. The evolution of the magnetic field
under the action of mass loss is left to a future study. The result is
presumably somewhat dependent on the initial choice for the current
distribution, and differences in the thermal evolution will have to be
taken into account.

It is important to understand the spreading of matter away from the
polar cap. This topic has received little attention, and is not
well-understood. Studies of the polar cap of accreting neutron stars
(Hameury et al.~1983; Brown \& Bildsten 1998; Litwin, Brown, \& Rosner
2001) and white dwarfs (Livio 1983; Hameury \& Lasota 1985) suggest
that spreading occurs once the sideways force due to the hydrostatic
overpressure overcomes the confining magnetic tension, or sooner if
instabilites play a role (Litwin et al.~2001). Understanding the
spreading process, and its affect on the magnetic field geometry in
the outermost layers of the star is directly relevant for interpreting
observations of polar cap fields of AM Hers. A related issue which
remains to be explored is the stability of the magnetic field
configurations found in \S 4. For example, stability considerations
demand a toroidal field component which is not included in our simple
models (see discussion in Mestel 1999).

Finally, we have not discussed the effect of white dwarf rotation. As
mentioned in the introduction, King (1985), applying the results of
Moss (1979) to accreting white dwarfs, suggested that meridional
currents could be responsible for the submergence of magnetic flux in
the outer layers of rotating magnetic white dwarfs. The aim was to
explain the narrow range of field strengths in AM Hers. The effect of
meridional flows in accreting white dwarfs deserves further
investigation.

\section*{Acknowledgments}

We thank Dayal Wickramasinghe for conversations about observations of
magnetic white dwarfs, and Graham Wynn for emphasising the lack of
very strongly-magnetic accreting white dwarfs. We are grateful to
Ellen Zweibel for discussions about the appropriate surface boundary
conditions in our global models, and to the referee Jean-Marie Hameury
for carefully questioning our assumptions and conclusions. We thank
Hugh van Horn and Alex Muslimov for communications regarding earlier
work, Phil Arras and Chris Thompson for discussions about the Hall
effect, and Dean Townsley and Lars Bildsten for discussions about
their models of compressional heating. We acknowledge support from
NASA through Hubble Fellowship grant HF-01138 awarded by the Space
Telescope Science Institute, which is operated by the Association of
Universities for Research in Astronomy, Inc., for NASA, under contract
NAS 5-26555.

\appendix

\section{Electrical Conductivity for Arbitrary Degeneracy}

In the white dwarf envelope, the electrons go from being
non-degenerate to degenerate. Hubbard \& Lampe (1969) calculated the
thermal conductivity for arbitrary degeneracy, presenting their
results in tabulated form. Unfortunately, other calculations of
conductivity which present their results in terms of fitting formulae,
such as those of Yakovlev \& Urpin (1980) and Itoh et al.~(1983) are
for degenerate electrons only (for thermal conductivity, this is the
regime of interest, since radiative heat transport dominates at low
densities). In their study of ohmic decay in cooling white dwarfs, WVS
adopted an interpolation scheme between calculations in the degenerate
and non-degenerate regimes. We have taken a different approach, which
we describe below. After our calculations were completed, we learned
of the new conductivity calculations by Potekhin (1999) and Potekhin
et al.~(1999) which treat the regime of intermediate
degeneracy\footnote{We thank Alexander Potekhin for bringing these
results to our attention. Details of these calculations and computer
codes to calculate conductivities can be found at {\tt
http://www.ioffe.rssi.ru/astro/conduct/index.html}}. Comparing with
these calculations, we find that our method agrees to within 20\% for
degenerate electrons and to within a factor of 2 for non-degenerate
electrons, adequate for our purposes.

For non-relativistic electrons, the density and electrical
conductivity may be written
\begin{equation}\label{eq:appne}
n_e={\sqrt{2}(m_ek_BT)^{3/2}\over \pi^2\hbar^3}F_{1/2}\left(\alpha\right)
\end{equation}
and
\begin{equation}\label{eq:appsig}
\sigma={2m_e(k_BT)^3\over\pi^3\hbar^3\Lambda_{ei}Zn_ee^2}
F_{2}\left(\alpha\right),
\end{equation}
where $\alpha=-(E_F-m_ec^2)/k_BT$, and the Fermi integrals $F_n(\eta)$
are given by
\begin{equation}
F_n\left(\alpha\right)=\int_0^\infty {x^ndx\over 1+\exp(x+\alpha)}.
\end{equation}
In equation (\ref{eq:appsig}), we assume the Coulomb logarithm
$\Lambda_{ei}$ varies slowly enough with electron energy that it may
be taken out of the integral. For degenerate electrons, $\alpha\ll
-1$, equation (\ref{eq:appsig}) reduces to the small $x$
(non-relativistic) limit of equation (\ref{eq:sig}). For
non-degenerate electrons, $\alpha\gg 1$, equation (\ref{eq:appsig})
reduces to equation (\ref{eq:spitzer}) (Spitzer 1962).

Our approach in this paper is to evaluate $\sigma$ using equation
(\ref{eq:appsig}), with $\Lambda_{ei}$ a function of density, and with
a correction for relativistic effects. We evaluate the Fermi integral
$F_2(\alpha)$ by direct numerical integration. The electron number
density is used to determine $\alpha$, by inverting equation
(\ref{eq:appne}) using the fit of Antia (1993) to the inverse of
$F_{1/2}$.

We calculate the Coulomb logarithm as a function of density by
interpolating between the non-degenerate result of Spitzer (1962) and
the degenerate results of Yakovlev \& Urpin (1980). We adopt the
expression of Yakovlev \& Urpin (1980) for the Coulomb logarithm, but
rewrite it in terms of the electron momentum, choosing $x$ in such a
way that in the non-degenerate limit, we obtain Spitzer's formula. The
Coulomb logarithm is $\Lambda_{ei}=\Lambda_{ei}^{(0)}-v^2/2c^2$, where
$\Lambda_{ei}^{(0)}=\ln(r_{\rm max}/r_{\rm min})$ (YU). Here $r_{\rm
min}$ and $r_{\rm max}$ are the limits of the integral over impact
parameters. For temperatures greater than $4.2\times 10^5\ {\rm K}$
(Spitzer 1962), $r_{\rm min}$ is set by the de~Broglie wavelength of
the electrons, we take $r_{\rm min}=\hbar/2p_e$. The cutoff $r_{\rm
max}$ is set either by the Debye length, given by $r_D^2=m_pk_BT/4\pi
e^2\rho\sum Y_iZ_i^2$, or the interion spacing $a=(3/4\pi n_i)^{1/3}$,
whichever is larger. We obtain
\begin{eqnarray}
\Lambda_{ei}^{(0)}&=&\ln\left[\left({2\over 9\pi}\right)^{1/3} {m_ec\over \hbar}
{x\over (\sum n_i)^{1/3}} \left({3\over\Gamma}+{3\over
2}\right)^{1/2}\right]\nonumber\\
&=&\ln\left[{127\ x\over \rho^{1/3}(\sum Y_i)^{1/3}}
\left({3\over\Gamma}+{3\over 2}\right)^{1/2}\right]
\end{eqnarray}
where
\begin{equation}
\Gamma={e^2\over k_BTa}{\sum n_i Z_i^2\over\sum n_i}=
0.11\left({\sum_iY_i Z_i^2\over\sum_iY_i}\right)
{\left(\rho_5\sum_iY_i\right)^{1/3}\over T_8},
\end{equation}
(Hubbard \& Lampe 1969; $\Gamma=(1/3)(a/r_D)^2$), and where
$x=(x_1^2+x_2^2)^{1/2}$, with
$x_2=(3k_BT/m_ec^2)^{1/2}=0.22\,T_8^{1/2}$.

Finally, we include a correction for relativistic effects. Because we
use equation (\ref{eq:appne}) to solve for $\alpha$, we obtain the
non-relativistic value for $\alpha$ in the degenerate limit, and
equation (\ref{eq:appsig}) reduces to the small $x$ limit of equation
(\ref{eq:sig}). To include relativistic effects, we divide equation
(\ref{eq:appsig}) by a factor $1+x^2$, so that we recover equation
(\ref{eq:sig}) in full in the degenerate limit.

For the $T=0$ models of \S 2 and \S 4, we write out
$\Lambda_{ei}^{(0)}$ as a sum of logarithmic terms, but drop the term
containing the temperature $T$, which contributes $<10$\% to the final
value of $\Lambda_{ei}$.

\end{document}